\documentclass[lettersize,journal]{IEEEtran}
\usepackage{amsmath,amsfonts,amssymb}
\usepackage{array}
\usepackage[caption=false,font=footnotesize,labelfont=rm,textfont=rm]{subfig}
\usepackage{textcomp}
\usepackage{xpatch}
\usepackage{diagbox}
\usepackage{url}
\usepackage{verbatim}
\usepackage{graphicx}
\usepackage{ragged2e}
\usepackage{cite}
\usepackage[table]{xcolor}
 \usepackage{threeparttable}
\usepackage{makecell}
\usepackage{hyperref}
\usepackage{amssymb}
\usepackage{multirow}
\usepackage{stmaryrd}
\usepackage{booktabs}
\usepackage{epstopdf}
\usepackage{wasysym}
\usepackage{pifont}
\usepackage[ruled,vlined]{algorithm2e}
\usepackage{algorithmic}
\usepackage{amsthm}
\usepackage{indentfirst}
\usepackage{soul}
 \usepackage{diagbox}
\SetKwFor{SSS}{Server:}{}{endfor}
\SetKwFor{BBB}{$S_2$:}{}{endfor}
\SetKwFor{clients}{Node:}{}{endfor}
\newtheorem{assumption}{Assumption}
\newtheorem{myDef}{Definition}
\newtheorem{Remark}{Remark}

\newtheorem{thm}{\bf Theorem}

\usepackage{color, xcolor}
\begin{document}
\title{Defending Against Backdoor Attacks via  Alignment Checking in Model-Contrastive Federated Learning}
\author{Hongliang Zhang,   Zhongyuan Yu, Guijuan Wang, Tianqing He, Wenshuo Ma, \\Xiaosong Zhang, Jiguo Yu,~\IEEEmembership{Fellow,~IEEE}

\thanks{This work was partially supported by NSF of China under Grants 62272256 and 62202250,  and the Shandong Province Youth Innovation Team Project under Grant 2024KJH032. ($\textit{Corresponding author}$: $\textit{Jiguo Yu}$)}
\thanks{H. Zhang, W. Ma, and G. Wang are with the Key Laboratory of Computing Power Network and Information Security, Ministry of Education, Shandong Computer Science Center, Qilu University of Technology (Shandong Academy of Sciences), Jinan, 250353, China, Email: b1043123004@stu.qlu.edu.cn, guijuan$\_$wang@126.com, weimws@foxmail.com.}
\thanks{Z. Yu is with the College of computer science and technology, China University of Petroleum,  Qingdao, 266580, China, Email:  yuzhy24601@gmail.com.}
\thanks{T. He, and X. Zhang are with  School of Computer Science and Engineering, University of Electronic Science and Technology of China, Chengdu, 611731, China, Email: sunny.he@std.uestc.edu.cn, johnsonzxs@uestc.edu.cn.}
\thanks{J. Yu is with  School of Computer Science and Engineering, University of Electronic Science and Technology of China, Chengdu, 611731, China, and also with the Big Data Institute, Qilu University of Technology, Jinan, 250353, China, Email: jiguoyu@sina.com; jiguoyu17@uestc.edu.cn.}

}

\markboth{Journal of \LaTeX\ Class Files,~Vol.~14, No.~8, August~2021}%
{Shell \MakeLowercase{\textit{et al.}}: A Sample Article Using IEEEtran.cls for IEEE Journals}


\maketitle

\begin{abstract}
 Federated Learning (FL)  is  vulnerable to backdoor attacks because of its distributed nature in edge computing scenarios.
Existing defense methods show limited efficacy as they overlook the   deviations among benign local updates caused by statistical  heterogeneity and the stealthiness of  backdoor attacks.
To tackle these issues, we propose  FedDAB, a  two-phase  method that combines local contrastive regularization with alignment  checking, to defend against backdoor attacks.
In the first phase,  FedDAB introduces a novel model-contrastive term into the local objective to  enhance direction and magnitude consistency among benign  updates.
In the second phase, FedDAB employs an alignment checking strategy  to evaluate each  local update  in terms of overall-direction alignment and  parameter-level alignment with historical information,   excluding updates that exhibit abnormal alignment patterns  from global  aggregation.
We theoretically prove FedDAB's robustness with a convergence rate of $\mathcal{O}(1/T)$.
 Extensive experiments show that FedDAB  outperforms   existing defense methods against  backdoor attacks.
\end{abstract}


\section{INTRODUCTION}

Edge Computing (EC) pushes  computation and storage resources from a centralized server  closer to edge nodes, thereby enabling tasks to be processed locally.
Building on this paradigm, federated learning allows multiple edge nodes to collaboratively train a high-performance global model without sharing their raw data \cite{chen2024federated}.
However, the distributed nature of FL inherently makes it   vulnerable to backdoor attacks in EC \cite{li2025backdoor}.
Such attacks maintain  the global model's accuracy on clean samples (i.e.,  main task) while inducing samples with specific features to be misclassified into an attacker-chosen target class (i.e.,  backdoor task). 
This vulnerability  stems from the server's inability to directly monitor each node's  data or training process.
Thus, an attacker may hijack a subset of nodes to poison their    data or manipulate their   training processes, causing them to submit malicious  updates that compromise the optimization of the global model.

To identify malicious  updates,  existing defense methods primarily focus on magnitude-based or direction-based strategies.
Specifically, the  methods in \cite{blanchard2017machine,huang2023multi,krauss2023mesas,11202428} employ  Euclidean distance to quantify the magnitude deviations among local updates,   whereas those in \cite{fung2020limitations,nguyen2022flame,10387506,10475552} adopt cosine similarity to capture their directional  relationships.
Nevertheless,  these methods  only capture   the overall information of each local update, while overlooking the parameter-level  details.
Concretely, since each local update is high-dimensional,   malicious nodes can  launch stealthy poisoning attacks by perturbing a small number of parameters (e.g., flipping  signs or amplifying  magnitudes) without noticeably changing  the overall direction or magnitude of their updates, thereby bypassing the aforementioned defense methods.
Furthermore, since data in FL  exhibit statistical heterogeneity, i.e., the data is Non-Independent and Identically Distributed (Non-IID), the effectiveness of these methods is further limited.
This is because  benign  updates  exhibit significant deviations   under Non-IID data \cite{li2020federated}\cite{pi2023dynafed}\cite{11075614}, thereby blurring the boundary between malicious and benign updates \cite{nguyen2022flame}\cite{11298358}\cite{11122662}.
Since each  update is inherently a vector  characterized by both direction and magnitude, the deviations among updates are also reflected in both aspects.
Thus, improving the consistency of benign  updates in both direction and magnitude  is crucial for defending against stealthy poisoning attacks in FL under Non-IID data.

To improve   update consistency,  several studies \cite{li2021model,son2021compare,10143270,seo2024relaxed,zhang2025swim,10705896} 
incorporate contrastive learning \cite{chen2020simple} into the local training objective as a regularization term,  thereby reducing the directional deviation  among local updates in FL.
However, these studies   overlook the magnitude deviation, resulting in residual inconsistency among benign updates.
This limitation motivates us to pose the   question: \textit{\textbf{Is there a solution that uses contrastive learning to enhance the   directional and magnitude consistency, while checking local updates from both  overall-direction  and  parameter-level  perspectives?}}

In this paper, we propose    FedDAB, a \underline{\textbf{Fed}}erated learning defense method designed to \underline{\textbf{D}}efend \underline{\textbf{A}}gainst \underline{\textbf{B}}ackdoor attacks.
The method integrates local contrastive regularization to enhance the consistency of local updates among benign nodes, and leverages alignment checking to  exclude suspicious updates from nodes.
Its  novelties   are as follows:
(\textbf{i}) A novel model-contrastive term is incorporated into the local optimization objective based on contrastive learning.
It regularizes benign local updates in both direction and magnitude, mitigating the inconsistency among benign updates caused by Non-IID data.
(\textbf{ii})  We examine local updates through both overall-direction checking and parameter-level  checking  to filter out  updates exhibiting abnormal alignment patterns.
(\textbf{iii})
We incorporate each node's historical behaviors into the parameter-level checking, instead of using only the  information from the current round.

Our  contributions are three-fold.
\begin{itemize}
  \item 
      To the best of our knowledge, FedDAB is the first FL defense method that combines local contrastive regularization with alignment checking, to defend  against backdoor attacks under varying data distributions. 
      
  \item We prove that FedDAB satisfies a convergence rate of $\mathcal{O}(1/T)$.  Furthermore, we provide a theoretical analysis of the robustness and  propagation error of FedDAB. 
      
  \item We conduct extensive experiments on multiple  datasets under Non-IID settings to evaluate the performance of FedDAB against various  backdoor attacks. 
      Compared to existing defense methods, FedDAB exhibits superior effectiveness and robustness.
\end{itemize}

\section{Related Works}
\textit{Federated Learning:} A typical FL system consists of a central server and $K$ nodes indexed by $k \in \mathcal{K}$, each of which holds its private training dataset  $\mathcal{D}_k$.
The nodes can cooperatively learn a global  model $W \in \mathbb{R}^d$ for the main task. 
The global optimization problem of FL can be formulated as:
 \begin{equation}\scalebox{1}{$
\min\limits_{W}F(W) =  \sum_{k\in \mathcal{K}} a_k F_k(W,\mathcal{D}_k),$} \nonumber
\end{equation}
where  \scalebox{0.89}{$a_k: = |\mathcal{D}_k|/ \sum_{k\in \mathcal{K}} |\mathcal{D}_k|$} is the data proportion of node $k$, and $F_k(\cdot)$ is the local objective function.
However, FL is  vulnerable to backdoor attacks, where an attacker can poison  the training data by embedding   backdoor triggers  into training samples  with specific  features.
To improve the attack stealthiness, the works in \cite{wang2020attack}\cite{zhang2022neurotoxin}   propose the variants of backdoor attacks that align malicious updates with historical global updates, making them less distinguishable from benign updates and enabling them to evade detection. 
These stealthy backdoor attacks further increase the difficulty of identifying malicious  updates in FL.

\textit{Defending against Backdoor Attacks:} 
Regardless of whether existing defense works are magnitude-based or direction-based, they generally mitigate the adverse impact of malicious nodes by assigning lower weights to suspicious updates or filtering them out before aggregation.
Specifically,   the works in \cite{fung2020limitations}\cite{10387506} \cite{9721118}\cite{ozdayi2021defending} retain   all local  updates  but assign lower aggregation weights to those identified as malicious, thereby limiting their influence on the global model.
For example, Yang et al.   observe  that updates from malicious nodes exhibit higher pairwise similarity than those from benign nodes \cite{10387506}.
Leveraging this property, they use adaptive clustering to group highly similar updates, aggregates updates within each cluster,  and  applies  projection-based dimensionality reduction to extract a plausible clean update, thereby reducing the contribution of malicious updates.
In contrast, the works in \cite{blanchard2017machine,huang2023multi,krauss2023mesas,10475552}  detect and remove malicious   updates before updating the global model.
Notably, the aforementioned rely on evaluating  the deviations among   updates in terms of  magnitude or direction.
However, FL inevitably suffers from statistical heterogeneity, i.e., Non-IID data, which leads to significant deviations  in both direction and magnitude among benign  updates \cite{ye2023heterogeneous}.
Thus, these defense works \cite{blanchard2017machine,huang2023multi,krauss2023mesas,fung2020limitations,10387506,10475552,9721118,ozdayi2021defending}  struggle to distinguish whether the deviations stem from Non-IID data or backdoor attacks, thereby  undermining their effectiveness.
Moreover, since backdoor attacks aim to maintain high accuracy on the main task while maximizing the success rate of the backdoor task, malicious updates are crafted to appear plausible in both magnitude and direction.
Thus,  the works in \cite{fung2020limitations,nguyen2022flame,10387506,10475552} provide limited robustness because they capture only overall-direction information of local updates while overlooking parameter-level details.
To address these limitations, our  FedDAB incorporates  both   overall-direction  and  parameter-level information into the  aggregation process to  detect malicious  updates.

\textit{Contrastive Learning in FL:} 
Contrastive learning is a self-supervised representation learning  that learns effective data representations by contrasting  positive and negative  pairs \cite{chen2020simple}.
Its core idea is to reduce the distance  between  the representations of different augmented views of  the same sample (i.e., positive pairs), and increase the distance between the representations of augmented views of different samples (i.e., negative pairs).
To improve FL performance  under Non-IID data,  Yang et al.    first  introduce  a model-contrastive term  into the FL objective to regularize local updates via contrastive learning \cite{li2021model}.
Subsequently, several variants based on  the work in \cite{li2021model} have been proposed \cite{10192272,yu2023multimodal,10413546}.
These works enhance the directional consistency among local updates while neglecting the magnitude deviation, leaving residual inconsistency among benign nodes.
To this end, we design a novel model-contrastive term within the local  objective to enhance both directional and magnitude consistency among benign updates, thereby making malicious updates easier to detect.

\section{The Detailed Design of FedDAB}
Since FedDAB combines local  regularization with alignment checking, the  design of both phases is presented.

\subsection{Local Contrastive Regularization}

Due to the statistical heterogeneity of data,  benign nodes optimize their local models toward their respective  optima \cite{li2020federated}\cite{pi2023dynafed}, resulting in deviations among benign local updates in both direction and magnitude.
These deviations cause   defense methods \cite{blanchard2017machine,huang2023multi,krauss2023mesas,10387506,10475552} struggle to  distinguish malicious updates from benign ones.
To eliminate the deviations caused by Non-IID data,    the works in \cite{li2021model,zhang2025swim,10192272,10413546}    incorporate contrastive learning \cite{chen2020simple} into the local objective to  improve the directional consistency among local updates.
However, they overlook the magnitude deviation among   updates.
Thus, we introduce a magnitude regularization into the model-contrastive term,  jointly aligning the direction and magnitude among benign   updates.
In the following, we present the network architecture  and local  objective.

\textbf{\textit{Network Architecture:}} 
The network consists of two components: an encoder and a  classifier.
 The  encoder   extracts a  representation vector  from an input, and the classifier maps this representation vector to a probability distribution over classes.
For simplicity,   for any model $W$, we use $\operatorname{Ext}_{W}(\cdot)$ to denote the encoder output produced by model $W$.

\textbf{\textit{Local Objective:}} 
In FedDAB, the local optimization function consists of  a classification term $\mathcal{L}_{\mathrm{Cla}}(\cdot)$ and a model-contrastive term $\mathcal{L}_{\mathrm{Con}}(\cdot)$.
The former is used to compute the main task loss,   while the latter is used to compute the contrastive loss between the feature representations learned by the local and global models.
Specifically, in the $t$-th round, for every input $x$, given the global model $W^t$, the current local model $w_k^t$, and the previous local model $w_k^{t-1}$, we extract the global representation vector \scalebox{0.82}{$z_g^t := \operatorname{Ext}_{W^t}(x)$}, the current local representation \scalebox{0.82}{$z_k^t : =\operatorname{Ext}_{w_k^t}(x)$}, and the previous local representation \scalebox{0.82}{$z_k^{t-1}: = \operatorname{Ext}_{w_k^{t-1}}(x)$}.
Since the global model provides better representations than  local models, we aim to decrease the distance between \scalebox{0.85}{$z_k^t$} and \scalebox{0.85}{$z_g^t$}, and increase the distance between \scalebox{0.82}{$z_k^t$} and \scalebox{0.82}{$z_k^{t-1}$}.
Thus, similar to works \cite{li2021model} \cite{zhang2025swim}, the  model-contrastive term  $\mathcal{L}_{\mathrm{Con}}(\cdot)$ is defined as:
 \begin{equation}\scalebox{0.7}{$
 \begin{aligned}
 &\mathcal{L}_{\mathrm{Con}}(W^{t},w^{t-1}_k,w^{t}_k,x) =\\&-\log \frac{\exp(\frac{\operatorname{sim}(z_g^t, z_k^t )}{q_1})}{\exp(\frac{\operatorname{sim}( z_g^t, z_k^t )}{q_1})+\exp(\frac{\operatorname{sim}(z_k^{t-1}, z_k^t)}{q_1})}-\log
\frac{
\exp(\frac{\| z_k^{t-1} - z_{k}^{t} \|_1}{q_2})
}{
\exp(\frac{\| z_g^t - z_{k}^{t} \|_1}{q_2}) +
\exp(\frac{\| z_k^{t-1} - z_k^{t} \|_1}{q_2})
},
\end{aligned} $} \nonumber
\end{equation}
where $\operatorname{sim}(\cdot)$ is the cosine similarity function, $\Vert \cdot \Vert_1$ is the $\ell_1$-norm,  and $q_1$ and $q_2$ are  used to rescale the distances.
In the above formula, the first term, termed the direction-alignment term, encourages the current local representation to match the global representation in direction, rather than the previous local representation.
The second term, termed the magnitude-alignment term,  enforces the current local representation to be closer to the global representation and pushes it away from the previous local representation in magnitude.
Combining  the terms $\mathcal{L}_{\mathrm{Cla}}(\cdot)$ and $\mathcal{L}_{\mathrm{Con}}(\cdot)$, the loss for an input sample $(x, y)$ is given by:
 \begin{equation}\label{12312214}\scalebox{1}{$
\mathcal{L}  = \mathcal{L}_{\mathrm{Cla}}(w^{t}_k,(x,y))+\mu \mathcal{L}_{\mathrm{Con}}(W^{t},w^{t-1}_k,w^{t}_k,x),\nonumber$}
\end{equation}
where $\mu$ is the trade-off parameter.
Thus, the  local objective of node $k$ is 
\scalebox{0.9}{$
\min_{w_k} \mathbb{E}_{(x,y)\sim\mathcal{D}_k} [\mathcal{L}]$}, which minimizes the classification loss while aligning the current local representation with the global representation.
After completing $E$ local SGD iterations,  benign node $k$ obtains $w^{t+1}_k$ and uploads the local update \scalebox{0.9}{$\Delta^{t}_k := w^{t+1}_k-  W^t$}  to the server.
In contrast, malicious nodes violate the predefined training rules and submit manipulated  updates to the server.


\subsection{Alignment Checking}
The alignment checking   phase consists of two key modules: \textit{Update Evaluation} and \textit{Update Aggregation}, and its core steps are  shown in Algorithm \ref{61410511}.
Concretely,   \textit{Update Evaluation} module is used to evaluate and filter the suspicious updates submitted by nodes.
  \textit{Update Aggregation} module is applied to compute the global model.
Each module is presented in  below.

\begin{algorithm}[t]\label{61410511}
   \begin{small}
    \caption{$\textbf {Alignment\_Checking}$}
    \label{961011}
    \LinesNumbered
    \KwIn {$\{\Delta_k^t\}_{k\in \mathcal{K}}, r, H, \lambda_{DSS}, \lambda_{SAS}$ $\quad \triangleright$ $\{\Delta_k^t\}_{k\in \mathcal{K}}$ is the local update set, $r$ is the  selection ratio, $H$  is  the buffer length, $\lambda_{DSS}$ and $\lambda_{SAS}$ are   thresholds.
}
    \KwOut {$W^{t+1}$} 
Initialize retained node  set $\mathcal{K}^t \leftarrow \emptyset$;\\
 // $\mathtt{Update  \ Evaluation}$ //\\
 $\{\alpha_k^t\}_{k\in \mathcal{K}} \leftarrow$ $\{\operatorname{Direction }\_\operatorname{Check}(\Delta_k^t)\}_{k\in \mathcal{K}};$\\
    $P^t \leftarrow$ $ \operatorname{Sgn} \left( \sum_{k \in  \mathcal{K}} \operatorname{Sgn} \big( \Delta^{t}_{k} \big) + \sum_{h=1}^{H} P^{t-h} \right);$\\
    $p_k^t  \leftarrow \operatorname{Sgn} (\operatorname{Sgn} ( \Delta_{k}^{t}) + \sum_{h=1}^{H}   p_{k}^{t-h} );$\\
   $\{\beta_k^t\}_{k\in \mathcal{K}} \leftarrow$ $\{\operatorname{Param}\_\operatorname{Check}(\Delta_k^t, P^t, p^t_k,r)\}_{k\in \mathcal{K}};$\\
   \For{\text{each node} $k  \in  \mathcal{K}$}{
   $m_{k,1}^t  \leftarrow \operatorname{MZ\_score}(\{\alpha_k^t\}_{k\in \mathcal{K}});$\\
   $ m_{k,2}^t  \leftarrow \operatorname{MZ\_score}(\{\beta_k^t\}_{k\in \mathcal{K}});$\\
   \If{$m_{k,1}^t  \textless \lambda_{DSS}$ \text{and} $m_{k,2}^t \textless \lambda_{SAS}$}{
   $\mathcal{K}^t  \leftarrow \mathcal{K}^t  \cup  \{k\}$;\\} 
   } 
  // $\mathtt{Update \ Aggregation}$ //\\
  $c \leftarrow\operatorname{Med}(\{\Vert\Delta_k^t\Vert_2\}_{k\in \mathcal{K}^t});$\\
  $\Delta_g^t \leftarrow \frac{1}{|\mathcal{K}^t|} \sum_{k \in \mathcal{K}^t} \left( \Delta_k^t \cdot \min \left\{ 1, \frac{c}{\|\Delta_k^t\|_2} \right\} \right);$\\
  $W^{t+1} \leftarrow W^{t} -\eta_g \Delta_g^t;$
    \end{small}
\end{algorithm}

\subsubsection{Update   Evaluation}
Existing  defense methods   mainly inspect local updates using  magnitude-based  metrics (e.g.,  Euclidean/Manhattan distance)  or   direction-based metrics    (e.g., cosine similarity).
As the global model tends to converge, the magnitude of local updates gradually decreases, making the magnitude gap between benign and malicious updates less pronounced.
Thus,  magnitude-based detection is   unreliable for identifying malicious updates.
To this end, we evaluate the alignment patterns of  local updates via overall-direction checking and parameter-level checking, and then filter out updates with abnormal patterns.
Therefore, the \textit{Update Evaluation} module consists of  \textit{Overall-Direction  Checking}, \textit{Parameter-Level Checking}, and \textit{Anomaly Score Detection}.
 
\textbf{\textit{Overall-Direction  Checking:} }
The local  objective of FedDAB is designed to enhance  the consistency of local updates among benign nodes in both direction and magnitude.
However, in typical backdoor attacks\cite{gu2017badnets}\cite{xie2019dba}, malicious  nodes  optimize both the main task and the backdoor task, 
which may cause their  updates to deviate in direction from   those of benign nodes \cite{xu2025identify}.
Thus, the overall-direction information of each local update is informative for identifying malicious updates.
Motivated by this insight, the \textit{Overall-Direction  Checking} step evaluates the Direction Similarity Score ($\operatorname{DSS}$) of each node based on its  update direction.
Formally, the $\operatorname{DSS}$ value $\alpha_k^t$ of local update $\Delta^{t}_k$ is defined  as: \begin{equation}\scalebox{1}{
$
\alpha_k^t := (1/(K-1))\sum_{j\in\mathcal{K}, j\neq k}S_{kj}^t,\nonumber
$}
\end{equation}
where \scalebox{0.85}{$  S_{kj}^t := \langle \Delta_k^t, \Delta_j^t \rangle/(\|\Delta_k^t\| \cdot \|\Delta_j^t\|)$}.
The $\operatorname{DSS}$ value  $\alpha_k^t$  captures the  average cosine similarity  between $\Delta^{t}_k$ and the other local updates, reflecting its overall-direction alignment.
Benign nodes typically have consistent $\operatorname{DSS}$ values, whereas malicious nodes exhibit $\operatorname{DSS}$ values that are markedly offset from those of benign nodes.
Notably, since $\operatorname{DSS}$ value is computed from update directions rather than magnitudes, its variability during federated training continues to be utilized  for identifying malicious updates.

\textbf{\textit{Parameter-Level  Checking:}}
In stealthy backdoor attacks \cite{wang2020attack}\cite{zhang2022neurotoxin}, malicious nodes  can manipulate the signs and magnitudes of certain parameters within their local updates  without changing the  update direction.
Although the $\operatorname{DSS}$ value  captures the  directional information, its detection ability is weakened under such attacks.
This motivates us to examine parameter-wise signs and magnitudes for fine-grained evaluation.
Notably,  the works in  \cite{ozdayi2021defending}\cite{xu2022byzantine}\cite{10398739} leverage    parameter-sign information to identify malicious updates.
Although  they utilize  the sign information,    parameters with very small magnitudes (i.e., unimportant parameters) may  weaken their defense effectiveness.
In addition, since benign nodes  honestly   follow  the FL protocol in each round, their historical behaviors can provide reliable evidence for detecting malicious updates.
However, the above defense methods only utilize  local updates  from the current round, ignoring the node's historical behaviors.
Thus,   the checking step   evaluates local updates by combining the signs and magnitudes of  important parameters from both the current and historical updates of each node.

To effectively resist  stealthy backdoor attacks, we examine sign alignment between each node's update and the global  sign vector at the coordinates with large absolute magnitudes.
The global sign vector represents the dominant sign at each coordinate across all   local updates.
Thus, the global  sign vector is defined as:
\begin{equation}\scalebox{1}{$
P^t = \operatorname{Sgn} ( \sum_{k \in \mathcal{K}} \operatorname{Sgn} ( \Delta_k^t ) ),
$} \nonumber
\end{equation}
where $\operatorname{Sgn}(\cdot)$ denotes the element-wise sign operator that returns $-1$, $0$, or $+1$ for negative, zero, and positive elements, respectively.
Since benign nodes  follow the  FL protocol across all rounds, incorporating historical information can better reflect their stable behavior.
Thus,    the server maintains a global sign buffer $\mathcal{S}_g$ that stores the most recent $H$ global sign vectors for computing $P^t$, which is reformulated as:
\begin{equation}\scalebox{1}{$
P^{t}  = \operatorname{Sgn} ( \sum_{k \in  \mathcal{K}} \operatorname{Sgn} ( \Delta^{t}_{k} ) + \sum_{h=1}^{H} P^{t-h} ),
$} \nonumber
\end{equation}
where $H$ denotes the length of  buffer $\mathcal{S}_g$.
Based on the global  sign vector $P^t$,  the sign alignment degree of each node can be evaluated.
To reduce interference from unimportant parameters, we evaluate the sign alignment  on the top-$r\%$ of coordinates ranked by absolute value within each local update, where $r$ denotes the selection ratio.
To identify these coordinates,  we use the indicator $\operatorname{Top}_r(\cdot)$ to mark them, which is defined as follows.

\begin{myDef}
(Indicator $\operatorname{Top}_r(\cdot)$).
Given a vector $Y \in \mathbb{R}^d$, and a selection ratio  $r \in [0,1]$, we define the function $\operatorname{Top}_r(\cdot):  \mathbb{R}^d \rightarrow  \{0,1\}^d$ as follows:
\begin{equation}
[\operatorname{Top}_r(Y)]_j =
\begin{cases}
1, & \text{if } |Y_j| \in \{|Y_{\pi(1)}|,\dots, |Y_{\pi(\lceil r\times d\rceil)}|\}, \\[2pt]
0, & \text{otherwise}, \nonumber
\end{cases}
\end{equation}
where $[\operatorname{Top}_r(Y)]_j$ denotes the $j$-th element of the indicator vector $\operatorname{Top}_r(Y)$, and $\pi(\cdot)$ is a permutation of the index set $\{1,2,\cdots,d\}$ such that:
\scalebox{0.8}{$
|Y_{\pi(1)}| \geq |Y_{\pi(2)}| \geq \cdots \geq |Y_{\pi(d)}|,
$}
where $\pi(d)$ is  the index of the element in vector  $Y$ that ranks the $d$-th largest in absolute value.
\end{myDef}

The indicator $\operatorname{Top}_r(\Delta^{t}_{k})$    outputs a binary  vector of the same size as the local  update $\Delta^{t}_{k}$, where each element is either 1 or 0.
The coordinates with a value 1 indicate the top-$r\%$ coordinates in $\Delta^{t}_{k}$ with the largest absolute magnitudes.
Using these coordinates, the  checking step measures each node's Sign Alignment Score ($\operatorname{SAS}$), which is defined as follows.
\begin{myDef}\label{26142017}
(Sign Alignment Score).
For vectors $Y_1, Y_2 \in \mathbb{R}^d$, the sign alignment score value $\beta$ of $Y_1$ and $Y_2$  is defined as:
\scalebox{0.85}{
$
\beta = 1 - \frac{\| \operatorname{Sgn}(Y_1) - \operatorname{Sgn}(Y_2) \|_0}{d},
$}
where $\| \cdot \|_0$ denotes the $\ell_0$-norm that counts non-zero elements. 
A larger $\beta$ means stronger coordinate-wise sign alignment between $Y_1$ and $Y_2$.
\end{myDef}


According to Definition \ref{26142017},   each node's $\operatorname{SAS}$ value $\beta_k^t$ is computed at the server by comparing its local sign vector $p^t_k$ with the global sign vector $P^t$.
Hence, the server constructs $p_k^t$ for each node before computing its $\operatorname{SAS}$ value.
Since $P^t$ incorporates historical information, each node's local sign vector likewise incorporates its own historical information.
 Thus, the server maintains a local sign buffer $\mathcal{S}_l$ that stores the most recent $H$ local sign vectors for each node, based on which $p_k^t$ is defined as:
\begin{equation}\scalebox{1}{$
p_k^t = \operatorname{Sgn} (\operatorname{Sgn} ( \Delta_{k}^{t}) + \sum_{h=1}^{H}   p_{k}^{t-h} ),
$} \nonumber
\end{equation}
 where $H$ denotes the length of  buffer $\mathcal{S}_l$,  set equal to that of buffer $\mathcal{S}_g$.
To focus on  the  coordinates with the largest absolute magnitudes,  we use the  $\operatorname{Top}_r(\cdot)$ indicator to   select the top-$r\%$ coordinates of each local  update. 
Thus,  the $\operatorname{SAS}$  value $\beta_k^t$ for node $k$ is computed  as follows:
\begin{equation}\scalebox{1}{$
\beta_k^t = 1 - \left\| \left(p_k^t - P^t \right) \odot \operatorname{Top}_r(\Delta_{k}^{t}) \right\|_0/\lceil r\times d\rceil,
$} \nonumber
\end{equation}
where   $\odot$ is the Hadamard product,  $\|\cdot\|_0$ is the $\ell_0$-norm, $(p_k^t- P^t)$  captures the coordinate-wise discrepancy between the local  and the global  sign vectors, and $\lceil r\times d\rceil$ (with $\lceil  \cdot\rceil$ being the ceiling operator) denotes the number of selected coordinates.
After computing each node's $\operatorname{SAS}$ value, the server stores the global sign vector $P^t$ in $\mathcal{S}_g$ and   each node's local sign vector $p_k^t$ in $\mathcal{S}_l$.
Once the buffer reaches its capacity, the oldest sign vector is replaced by the newly computed one.

\textbf{\textit{Anomaly Score Detection:}}
After  \textit{Overall Direction Checking} and \textit{Parameter-Level Checking}, 
a  Median-based Z-score  is introduced to   remove the local updates with abnormal   $\operatorname{DSS}$  and $\operatorname{SAS}$ values.
 The \textit{Anomaly Score Detection} step is detailed below.

Since malicious nodes perturb their local updates  to launch backdoor tasks, their  $\operatorname{DSS}$  and $\operatorname{SAS}$ values exhibit noticeable offsets from those of benign nodes.
Assuming that most nodes are benign, nodes whose $\operatorname{DSS}$  and $\operatorname{SAS}$ values lie close to the medians of $\{\alpha_k^t\}_{k \in \mathcal{K}}$ and $\{\beta_k^t\}_{k \in \mathcal{K}}$ are regarded as benign, while those that are far from the medians are considered suspicious.
Thus, the median  offsets  of node $k$ are defined as
\scalebox{0.84}{$m_{k,1}^t :=|\alpha_k^t- \operatorname{Med}(\{\alpha_k^t\}_{k\in \mathcal{K}})|,$} and 
\scalebox{0.84}{$m_{k,2}^t :=|\beta_k^t- \operatorname{Med}(\{\beta_k^t\}_{k\in \mathcal{K}})|,$}
where $\operatorname{Med}(\cdot)$ is the median operator.
We use the thresholds $\lambda_{DSS}$ and $\lambda_{SAS}$ to identify nodes whose median offsets exceed these bounds.
However, since  the ranges of median offsets  vary substantially across rounds,  fixed  thresholds  $\lambda_{DSS}$ and $\lambda_{SAS}$ may cause all nodes' offsets to fall within these bounds, thereby failing to filter out anomalous updates.
To address this, inspired by the  standardization method Z-score, we introduce  a robust variant named  median-based Z-score $\operatorname{MZ\_score}(\cdot)$ to recalculate  the median offsets,  adapting to changes in the  $\operatorname{DSS}$  and $\operatorname{SAS}$  ranges during federated training.
\begin{myDef}\label{7231752}
(Median-based Z-score $\operatorname{MZ\_score}(\cdot)$). Let $X$ be a set of scalars with median $\operatorname{Med}(X)$ and standard deviation $\operatorname{Std}(X)$.
For any    $s \in X$, its   offset   is defined by:
\scalebox{0.9}{$
\operatorname{MZ\_score}(s, X)=  (\left|s - \operatorname{Med}(X)\right|)/\operatorname{Std}(X),  
$}
which measures the standardized distance of scalar $s$ from   $\operatorname{Med}(X)$.
\end{myDef}

By applying   $\operatorname{MZ\_score}(\cdot)$ function to the  sets \scalebox{0.84}{$\{\alpha_k^t\}_{k \in \mathcal{K}}$} and  \scalebox{0.84}{$\{\beta_k^t\}_{k \in \mathcal{K}}$}, 
we recompute each node's median offsets \scalebox{0.84}{$m_{k,1}^t$} and \scalebox{0.84}{$m_{k,2}^t$}.
Based on these  offsets, nodes whose offsets exceed the thresholds are filtered out.
Specifically, if  \scalebox{0.87}{$m_{k,1}^t \geq \lambda_{DSS}$} or \scalebox{0.87}{$m_{k,2}^t \geq \lambda_{SAS}$}, its   update \scalebox{0.87}{$\Delta_k^t$}  is identified as suspicious, and node $k$ is excluded from aggregation in the $t$-th round.
Thus, the retained node set $\mathcal{K}^t$  is  defined as:
\begin{equation}\scalebox{1}{$
\mathcal{K}^t = \{k \in \mathcal{K}|(m_{k,1}^t \textless \lambda_{DSS}) \land (m_{k,2}^t \textless \lambda_{SAS}) \}.$} \nonumber
\end{equation}
Notably, $\lambda_{DSS}$ and $\lambda_{SAS}$ are   fixed thresholds.
The server aggregates updates from \scalebox{0.84}{$\mathcal{K}^t$}  to compute the global model.

\subsubsection{Update Aggregation} 
After filtering, the remaining nodes are  treated as benign, and their updates are used to calculate the global update $\Delta_g^t$.
However,    the \textit{Update Evaluation} module is unable to detect malicious updates with abnormally large magnitudes.
If update magnitudes are not considered,  malicious updates may bypass \textit{Update Evaluation},  enabling malicious nodes to compromise the global model via amplifying update magnitudes.
To this end, FedDAB employs the median  $\ell_2$-norm of the retained updates in $\mathcal{K}^t$ as the clipping threshold,    rescaling these  updates.
Formally, the  aggregation process is expressed as:
\begin{equation}\scalebox{1}{$
\Delta_g^t = \frac{1}{|\mathcal{K}^t|} \sum_{k \in \mathcal{K}^t} \left( \Delta_k^t \cdot \min \left\{ 1, \frac{c}{\|\Delta_k^t\|_2} \right\} \right),\label{1172236}
$} \nonumber
\end{equation}
where $c$ denotes the clipping bound, defined as
\scalebox{0.9}{$
c :=\operatorname{Med}(\{\Vert\Delta_k^t\Vert_2\}_{k\in \mathcal{K}^t}).
$}
The global model \scalebox{0.9}{$W^t$} is  updated   as
\scalebox{0.9}{$
W^{t+1} := W^{t} -\eta_g \Delta_g^t,
$}
where $\eta_g$ is the global learning rate.

\section{Robustness Analysis of FedDAB}
This section provides the theoretical analysis of  FedDAB.
Before presenting the theoretical results, we state the following assumptions.
Note that Assumption \ref{101231345} and \ref{10121537} have been widely adopted  in the theoretical analysis of FL \cite{xu2022byzantine}\cite{11298519}.
Assumption \ref{101231346} provides a standard measure for the deviations caused by data heterogeneity among nodes \cite{xu2025detecting} \cite{allouah2023fixing}.
Such heterogeneity complicates the problem of FL with backdoor attacks, as it may cause  the server to confuse malicious updates with benign ones.
The proofs of all theorems  are provided in Appendix.

\begin{assumption}\label{101231345}
($L_1$-smoothness). Let $\mathcal{K}_{\mathcal{B}}$ be the set of benign nodes. 
For each benign node $k \in \mathcal{K}_{\mathcal{B}}$, its  objective function $F_k(\cdot)$ is assumed to be $L_1$-Lipschitz smooth with $L_1 \textgreater 0$.
Formally, for any $Y_1, Y_2 \in \mathbb{R}^d$, the following inequality holds:
\begin{equation} \scalebox{1}{$
 \|\nabla F_k(Y_1) - \nabla F_k(Y_2)\|_2 \leq L_1 \|Y_1 - Y_2\|_2. \nonumber$}
\end{equation}
Moreover, this condition   implies the following inequality:
\begin{equation} \scalebox{1}{$
F_k(Y_1) - F_k(Y_2) \leq 
\nabla F_k(Y_1)^{\top}(Y_2 - Y_1) + \frac{L_1}{2}\|Y_1 - Y_2\|^2_2.\nonumber$}
\end{equation}
\end{assumption}

\begin{assumption} \label{10121537}
(Bounded  Gradient and  Variance).
For each benign node $k \in \mathcal{K}_{\mathcal{B}}$, the stochastic gradient $g_k$ is an unbiased and bounded-variance estimator of the  gradient. Specifically, for any $Y \in \mathbb{R}^d$, it holds that $\mathbb{E}[g_k] = \nabla F_k(Y)$, $\Vert\nabla F_k(Y) \Vert^2_2 \leq Z^2$, and $\mathbb{E}\|g_k - \nabla F_k(Y)\|^2_2 \leq \xi_k^2$ for all benign nodes, where the expectation is taken over local mini-batches. For convenience, the average variance bound is denoted   as $\bar{\xi}:= \frac{1}{|\mathcal{K}_{\mathcal{B}}|} \sum_{k \in \mathcal{K}_{\mathcal{B}}} \xi_k^2$.
\end{assumption}

\begin{assumption}\label{101231346}
(Bounded Heterogeneity).
For a constant $L_2$$\geq$$0$, it holds that \scalebox{0.8}{$\frac{1}{|\mathcal{K}_{\mathcal{B}}|} \sum_{k \in \mathcal{K}_{\mathcal{B}}} \|\nabla F_k(Y) - \nabla F_{\mathcal{K}_{\mathcal{B}}}(Y)\|^2_2 \leq L_2$} for all \scalebox{0.85}{$Y$$\in$$\mathbb{R}^d$}, where the mean gradient over the benign node set is defined as  \scalebox{0.85}{$\nabla F_{\mathcal{K}_{\mathcal{B}}}(Y):= \frac{1}{|\mathcal{K}_{\mathcal{B}}|} \sum_{k \in \mathcal{K}_{\mathcal{B}}} \nabla F_k(Y)$}.
\end{assumption}

Beyond the above assumptions, to facilitate the theoretical analysis of FedDAB, we define the robustness concept, namely $\kappa$-secure aggregation, as presented in Definition \ref{10051653}.
The definition is commonly adopted in the analysis of FL defense methods \cite{allouah2023fixing}\cite{allouah2023robust}.
\begin{myDef}\label{10051653}
($\kappa$-Secure Aggregation) Given the set of model updates $\{\Delta_k\}_{k\in \mathcal{K}}$ and benign node set $\mathcal{K}_{\mathcal{B}} \subset \mathcal{K}$, the aggregation rule $G(\cdot)$: 
 $\mathbb{R}^{d \times  K}   \rightarrow  \mathbb{R}^{d}$  is called $\kappa$-secure aggregation when its aggregated global update $\Delta_g: = G(\{\Delta_k\}_{k\in \mathcal{K}})$ satisfies 
$\Vert\Delta_g - \Delta_{\mathcal{K}_{\mathcal{B}}} \Vert^2_2 \leq \kappa$, where $\Delta_{\mathcal{K}_{\mathcal{B}}}: = (1/|\mathcal{K}_{\mathcal{B}}|)\sum_{k\in \mathcal{K}_{\mathcal{B}}}\Delta_k$, $K$ is the number of nodes, and $\kappa \geq 0$ denotes the robustness coefficient of $G(\cdot)$.
\end{myDef}

This definition quantifies a constant bound, denoted by $\kappa$, that constrains  the deviation between the global update output by  FedDAB and the average of all benign  updates.
A smaller $\kappa$ indicates that the aggregation function $G(\cdot)$ produces the global  update closer to the average of benign updates, which represents the optimal output.
In other words, when $G(\cdot)$  identifies and removes all malicious  updates while preserving all benign ones, we have $\kappa =0$, indicating that FedDAB achieves the highest level of $\kappa$-secure aggregation.


\begin{thm}[Bounded Local  Deviation] \label{local_divergence1231} In the $t$-th round,
let \scalebox{0.8}{$\Delta_{\mathcal{K}_\mathcal{B}}^t := (1/|\mathcal{K}_\mathcal{B}|) \sum_{k\in \mathcal{K}_\mathcal{B}}\Delta_k^t$} be the average of benign  updates.
If  local learning rate $\eta_l$  and  local iterations $E$ satisfy  $\eta_l\leq\frac{1}{2E}$, under   Assumption \ref{10121537} and  \ref{101231346}, the local deviation   is bounded as:
    \begin{equation}\scalebox{1}{$
        \frac{1}{|\mathcal{K}_\mathcal{B}|} \sum_{k \in \mathcal{K}_\mathcal{B}} \mathbb{E}\left\|  {\Delta}_k^t - \Delta_{\mathcal{K}_\mathcal{B}}^t \right\|^2_2 \leq 2\bar{\xi} + L_2. \nonumber$}
    \end{equation}
\end{thm}
\begin{Remark}
When the condition $\eta_l \leq 1/(2E)$ holds, mitigating the deviation $L_2$ induced by data heterogeneity can  constrain the local deviation $(2\bar{\xi} + L_2)$.
Thus, in FedDAB, the model-contrastive term is incorporated into the local  objective to align both the magnitude and direction of benign  updates,  thereby reducing the deviation  $L_2$.
\end{Remark}
\begin{thm}
[$\kappa$-robustness of FedDAB] \label{lemma:AlignInskappa4221}
Let Assumption \ref{10121537} and Assumption \ref{101231346} hold. 
Suppose that the total number of nodes is $K> 1$, and that there are $K_m$ malicious nodes satisfying $ 0 \leq K_m < K/(3+\epsilon)$ for a positive constant $\epsilon$.
In the $t$-th round, if the  relations $\eta_l\leq \frac{1}{2E}$ and $|\mathcal{K}^t| \geq K-2K_m$ holds, FedDAB achieves $\kappa$-secure aggregation, where $\kappa$ is defined by:   
\begin{equation}\scalebox{1}{$
\begin{aligned}
\kappa & = \left( 1 + K_m / \left( K-2K_m \right) \right) \left( \left(2/\epsilon + 1\right) \left(2\bar{\xi} + L_2\right) + 8 c^2 \right) \\
     &= \mathcal{O}\left( 1 + K_m / \left( K-2K_m \right)  \right). \label{ourkappa_order}
\end{aligned} \nonumber$}
\end{equation}
\end{thm}
\begin{Remark}
The relation  emphasizes the importance of preserving a  sufficient number of remaining nodes.
Moreover, the  clipping bound $c$ can mitigate the deviation caused by overly large-magnitude local updates.
In addition,  \cite{xu2022byzantine}\cite{panda2022sparsefed} have demonstrated the effectiveness of the clipping in mitigating the impact of malicious  updates.
We argue that FedDAB achieves comparable robustness to   defense methods, i.e., RFA \cite{9721118} $(\mathcal{O}(1+K_m / (K-2K_m))^2)$ and Krum \cite{blanchard2017machine} ($\mathcal{O}( 1 + K_m / ( K-2K_m ) )$).
The results for RFA and Krum are taken from \cite{allouah2023fixing}. 
Since the coefficient $\kappa$ is bounded by a constant, we  incorporate these results into our discussion without losing generality.
\end{Remark}

\begin{thm}[Propagation Error Bound] \label{lemma:AlignInscertified_radius}
Let Assumption \ref{101231345} to Assumption   \ref{101231346} hold, and let the local learning rate satisfying  $\eta_l \leq 1/(2E)$.  Let $W^T$ denote the  global model trained over all nodes in the $T$-th  round, and $W^{T,*}$ denote the clean global model trained only on  benign node set $\mathcal{K}_\mathcal{B}$. Under backdoor attacks,   the propagation error of FedDAB is bounded by:
    \begin{equation}\scalebox{1}{$
    \begin{aligned}
         \left \| W^{T} - W^{T, *} \right\|^2 \leq 4\tau\eta_g(\kappa+2\bar{\xi})((1+1/\tau)^T-1),
    \end{aligned} \nonumber$}
    \end{equation}
  where  $\kappa$ is given in Theorem  \ref{lemma:AlignInskappa4221},  and   $\tau := 1/(1+3\eta^2_gL_1^2)$.    
    \end{thm}
\begin{Remark}
The error bound  reflects the deviation between the clean  model $W^{T, *}$ and the actual model  $W^{T}$.
When the global learning rate $\eta_g$ is fixed, this error bound increases as the number of rounds $T$ grows.
To  mitigate the propagation error, the global learning rate is scheduled to decay over rounds, i.e., $\eta_g^t$  gradually decreases as  $t$ increases.
Thus, FedDAB can be integrated with  decaying global learning rate schedule \cite{wang2023fedhyper}  to achieve stable convergence.
\end{Remark}

     \begin{thm}
[Convergence Rate] Suppose Assumptions \ref{101231345} and \ref{10121537} hold.
Let   $W^\star$ denote the global optimum.
After $T$  rounds of   FedDAB, the following bound holds:
    \begin{equation}\scalebox{1}{$
\frac{1}{T}\sum_{t=1}^{T}\mathbb{E}\left\Vert\nabla F\left(W^t\right)\right\Vert^2_2
\leq \frac{4|\mathcal{K}^t|}{T\eta_g}\Phi +4H_2-4H_1 +\frac{2|\mathcal{K}^t|L_1 H_1}{\eta_g}, \nonumber$}
\end{equation}  
where    \scalebox{0.85}{$H_1 = (\eta_lEZ |\mathcal{K}^t|)^2$}, \scalebox{0.85}{$H_2 =  2Z^2 \left(\frac{1}{2}+|\mathcal{K}^t|\right)^2 + 4Z^2|\mathcal{K}^t|^2+4H_1$},   \scalebox{0.9}{$\Phi = F(W^0) - F(W^\star)$}, and $\eta_l$ is the local learning rate.
\end{thm}
\begin{Remark}
The bound indicates that the gradient norm of FedDAB  decreases as the number of rounds $T$ increases.
If all other parameters are fixed with respect to $T$,  the convergence rate of FedDAB  is $\mathcal{O}(1/T)$.
\end{Remark}

\section{Experimental Details}

\subsection{Experimental Settings}
\subsubsection{Hyper-parameter Configuration}
Similar to existing FL works  \cite{11202428}\cite{10387506}\cite{10704054},
we evaluate the defense methods on  FMNIST \cite{xiao2017fashion}, CIFAR10 \cite{krizhevsky2009learning}, and CIFAR100 \cite{krizhevsky2009learning}.
We simulate a FL system with 20 edge nodes, where all nodes participate in each training round.
Notably,  we further evaluate  the scalability of FedDAB under  larger-scale node settings in Section \ref{561449}.
Unless otherwise specified, we use the following default settings.
The  thresholds $\lambda_{DSS}$ and $\lambda_{SAS}$ are set to $1.0$.
The trade-off parameter $\mu$  is set to 0.5, and the buffer length $H$ for both $\mathcal{S}_l$ and $\mathcal{S}_g$ is set to 3.
The selection  ratio  $r$  is set as 30\%, meaning that the Top-30\% of model parameters are used for the parameter-level checking.
The global learning rate $\eta_g$ is set to $1$, the local learning rate $\eta_l$ to $0.05$, the number of local iterations $E$ to $2$, and the temperature parameters $q_1$ and $q_2$ to $1$. The number of  rounds $T$ is set to $50$ for FMNIST, $100$ for CIFAR10, and $120$ for CIFAR100.
For the model architectures, we use ResNet9 for FMNIST and CIFAR10, and VGG9 for CIFAR100.
To simulate the Non-IID setting, we partition each dataset among nodes using a Dirichlet distribution $\operatorname{Dir}(\Psi)$, where a larger concentration parameter $\Psi$ indicates a more uniform class allocation across nodes.

\subsubsection{Evaluation Metrics}
We use the following  metrics to evaluate the defense methods.
Test Accuracy (TA) is the proportion of clean  samples correctly classified by the global model.
Attack Success Rate (ASR) is the proportion of triggered  samples misclassified into the target label.
Robustness Rate (RR) is the proportion of triggered  samples correctly classified into their true labels.
An effective  defense method should achieve high TA and RR, while maintaining a low ASR.

\subsubsection{Attack Settings}
The attacker is assumed to control a proportion $v$ (default 30\%) of all nodes as malicious nodes.  
These malicious nodes poison their local training data by embedding a ``plus'' trigger into clean images.
Typical backdoor attacks include BadNet \cite{gu2017badnets} and Distributed Backdoor Attack (DBA) \cite{xie2019dba}. To improve attack stealthiness, more advanced attacks such as Neurotoxin attacks \cite{zhang2022neurotoxin}  and Projected Gradient Descent (PGD)-based attacks \cite{wang2020attack} have been further proposed.
      \begin{enumerate} 
        \item \textit{BadNet Attacks}.  Malicious nodes add uniform ``plus'' triggers  to clean samples  to construct  poisoned data.
        \item \textit{DBA Attacks}. The ``plus'' trigger is decomposed into  four sub-triggers, and each malicious node uses only one of these sub-triggers.
        \item \textit{Neurotoxin Attacks}.    Malicious nodes project  their local updates onto the coordinate set defined by the top $75\%$  largest-magnitude parameters of the previous global model, and sets the update in  other coordinates to zero.
        \item \textit{PGD Attacks}. The malicious local update is projected onto a sphere to form the tampered   update, where the sphere's radius is set to the $\ell_2$-norm of the previous-round global model.
      \end{enumerate}

\subsubsection{Evaluated Defense Methods}
We  compare our FedDAB   with the non-robust  FedAvg \cite{mcmahan2017communication} and eight defense methods:  MKrum \cite{blanchard2017machine}, Foolsgold \cite{fung2020limitations}, RLR \cite{ozdayi2021defending}, RFA \cite{9721118},  MMetric \cite{huang2023multi}, RoseAgg \cite{10387506},  AlignIns \cite{xu2025detecting},  and EndPCA \cite{11358404}.

\subsection{Experimental Results}

\begin{table*}[htbp]
  \centering
  \caption{The  ASR and RR results of baselines and FedDAB on the $\operatorname{Dir}(1.0)$ setting of FMNIST, CIFAR10, and CIFAR100.  The best and second-best results are shown in \textbf{bold} and \underline{underlined}, respectively.}
  \label{tab:main_results}
  \renewcommand{\arraystretch}{1}
  \scalebox{0.68}{
  \tabcolsep=0.065cm
    \begin{tabular}{c|c|cccccc|cccccc|cccccc|cccccc|cc}
    \toprule\midrule
    \multirow{3}[6]{*}{\textbf{\makecell*[c]{Dataset \\ (Model)}}} & \multirow{3}[6]{*}{\textbf{Methods}}  & \multicolumn{6}{c|}{\textbf{BadNet}}     & \multicolumn{6}{c|}{\textbf{DBA}}     & \multicolumn{6}{c|}{\textbf{Neurotoxin}}     & \multicolumn{6}{c|}{\textbf{PGD}} & \multirow{3}[6]{*}{\textbf{\makecell*[c]{Avg. \\ ASR$\downarrow$ }}} & \multirow{3}[6]{*}{\textbf{\makecell*[c]{Avg. \\ RR$\uparrow$ }}}\\
\cmidrule(r){3-8} \cmidrule(r){9-14} \cmidrule(r){15-20} \cmidrule(r){21-26}                &       & \multicolumn{3}{c}{ASR(\%)$\downarrow$} & \multicolumn{3}{c|}{RR(\%)$\uparrow$} & \multicolumn{3}{c}{ASR(\%)$\downarrow$} & \multicolumn{3}{c|}{RR(\%)$\uparrow$} & \multicolumn{3}{c}{ASR(\%)$\downarrow$} & \multicolumn{3}{c|}{RR(\%)$\uparrow$}& \multicolumn{3}{c}{ASR(\%)$\downarrow$} & \multicolumn{3}{c|}{RR(\%)$\uparrow$} \\
\cmidrule(r){3-5} \cmidrule(r){6-8} \cmidrule(r){9-11} \cmidrule(r){12-14} \cmidrule(r){15-17}  \cmidrule(r){18-20}   \cmidrule(r){21-23}  \cmidrule(r){24-26}      &             & $v$=0.30 & $v$=0.40& $v$=0.50 & $v$=0.30 & $v$=0.40& $v$=0.50 & $v$=0.30 & $v$=0.40& $v$=0.50 & $v$=0.30 & $v$=0.40& $v$=0.50 & $v$=0.30& $v$=0.40 & $v$=0.50 & $v$=0.30 & $v$=0.40& $v$=0.50 & $v$=0.30 & $v$=0.40& $v$=0.50 & $v$=0.30 & $v$=0.40& $v$=0.50  \\\midrule
        \multirow{8}[4]{*}{\rotatebox{90}{\makecell*[c]{FMNIST \\ (ResNet9)}}} & FedAvg & 92.38 &95.28& 98.32 & 6.84&4.26 &1.56  &74.14 & 22.52 &61.06 &21.74  &52.67 &29.46 &77.07&83.01& 88.30&20.22 &15.39 &10.71 &92.38& 95.28& 98.32&6.84  &4.26& 1.56 &81.50 & 14.62 \\ \cmidrule{2-28}   
                  & MKrum & \textbf{2.82} &5.46& 10.50 & \textbf{71.49} &56.20 & 45.95  &\textbf{2.24}& \underline{3.49}& 12.54&\textbf{68.22}  &59.38&40.90 &2.82&\textbf{2.78}&12.21 &\underline{71.49} &57.04 & 46.51 & \textbf{2.80}& 5.46& 13.08 & \textbf{71.53} &56.20& 37.90 & 6.35 & 56.89 \\
          & Foolsgold  &\underline{2.90} &\underline{3.18}& \textbf{2.96} & 65.79&67.86& 65.49&6.44 &8.25 &8.42 & 63.32 & 61.81 &\underline{60.41} &\underline{2.12} &3.52&\underline{6.54}&68.46 &62.35 &\textbf{76.42} &   4.53 &4.63& \underline{6.26} & 66.34& \textbf{72.63}& \textbf{77.47} & 4.97 &\textbf{67.36} \\     
          &    RLR  & 64.91 &51.76& 30.50 &22.56 &21.36& 14.80 &54.13 &60.67 & 62.96 & 20.06 &18.53& 15.82 &\textbf{0.73}&21.54&99.87& \textbf{72.99}&10.41 &0.11 &64.91 & 45.67& 30.50& 22.56 &21.36& 14.80& 49.01 & 21.28 \\
          & RFA   & 36.41 &41.07& 45.60  &49.77&47.53 &40.63 &5.31 & 14.83& 34.51 & 60.19 &52.51&37.87 &3.73&3.13& 9.25& 64.36& 71.74& 59.51 & 36.41&41.07 &48.51   & 49.77 &47.53& 37.49 & 26.65  & 51.57 \\
          &     MMetric & 48.13 &82.20& 84.91 &40.67 &14.77& 13.90 & 19.62& 11.40& 91.44 & 50.34 &45.43&7.67 &91.44&71.99&84.13 & 7.67&24.19 &14.33  &48.13 & 82.20& 84.91 & 40.67&14.77&13.90 & 66.70 &24.02 \\
&     RoseAgg &91.43 &95.47&96.31  &7.62 &4.00& 2.14 & 69.73& 39.86& 50.49 & 25.01 &40.48 & 29.14 &74.61 &81.44 & 89.20 & 22.30& 16.63 & 7.06  &41.80  & 54.64&  59.54 & 56.94  & 42.67&  36.50 & 70.37  & 24.20 \\
          &  AlignIns &2.68 &4.03& 8.44 & 68.11 &68.50 & 50.84  & \underline{2.83}& 3.79& \underline{6.73} & \underline{67.21} &\textbf{68.12}&47.69 &2.78&3.53&7.49 &66.83 &\underline{68.50} &53.18  & 3.23& \underline{4.52}& 7.98& 66.11 &\underline{67.50}& 54.04&\underline{4.83}  & 62.21 \\
                    &  EndPCA &2.03 &5.64& 8.24 & 65.42 &62.40 & 57.63  &3.88& 5.43&7.75 & 63.67 &61.09&58.33 &2.16&4.12&9.87 &66.49 &63.34 &57.51 & 3.92& 6.89&7.79&65.18 &62.93& 59.89&5.64  & 61.98 \\
          & \cellcolor[rgb]{ .816,  .808,  .808}FedDAB & \cellcolor[rgb]{ .816,  .808,  .808}2.91&\cellcolor[rgb]{ .816,  .808,  .808}\textbf{2.56}& \cellcolor[rgb]{ .816,  .808,  .808}\underline{4.17}&\cellcolor[rgb]{ .816,  .808,  .808}\underline{71.01} &\cellcolor[rgb]{ .816,  .808,  .808}\textbf{69.02} &\cellcolor[rgb]{ .816,  .808,  .808}\textbf{63.68}  &\cellcolor[rgb]{ .816,  .808,  .808}3.44 &\cellcolor[rgb]{ .816,  .808,  .808}\textbf{2.91} & \cellcolor[rgb]{ .816,  .808,  .808}\textbf{3.22}& \cellcolor[rgb]{ .816,  .808,  .808}66.53&\cellcolor[rgb]{ .816,  .808,  .808}\underline{66.79}&\cellcolor[rgb]{ .816,  .808,  .808}\textbf{64.13} &\cellcolor[rgb]{ .816,  .808,  .808}3.58&\cellcolor[rgb]{ .816,  .808,  .808}\underline{2.86}&\cellcolor[rgb]{ .816,  .808,  .808}\textbf{3.44} & \cellcolor[rgb]{ .816,  .808,  .808}66.22&\cellcolor[rgb]{ .816,  .808,  .808}\textbf{68.58} & \cellcolor[rgb]{ .816,  .808,  .808}\underline{68.20} & \cellcolor[rgb]{ .816,  .808,  .808}\underline{2.88}& \cellcolor[rgb]{ .816,  .808,  .808}\textbf{2.46}& \cellcolor[rgb]{ .816,  .808,  .808}\textbf{3.62} & \cellcolor[rgb]{ .816,  .808,  .808}\underline{66.88} &\cellcolor[rgb]{ .816,  .808,  .808}62.43&\cellcolor[rgb]{ .816,  .808,  .808}\underline{58.83} &\cellcolor[rgb]{ .816,  .808,  .808}\textbf{3.17} &\cellcolor[rgb]{ .816,  .808,  .808}\underline{66.02} \\ 
    \midrule
        \multirow{8}[4]{*}{\rotatebox{90}{\makecell*[c]{CIFAR10 \\ (ResNet9)}}} & FedAvg & 99.96 &99.94& 100 & 0.02&0.02 &0.00  &91.58 & 78.99 &98.31 &7.18  &19.52 &1.62 &99.82&99.67& 99.96&0.17 &0.27 &0.04 & 99.96& 99.94& 100&0.02  &0.02& 0.00  & 97.34 & 2.40 \\ \cmidrule{2-28}   
                  & MKrum & 99.77 &99.98& 100 & 0.21 &0.01 & 0.00  &85.73& 97.87& 100&10.53  &1.44&0.00 &98.73&99.99&100 &0.89 &0.01 & 0.00 & 99.84& 99.98& 100 & 0.17 &0.00& 0.00 & 98.49 & 1.10 \\
          & Foolsgold  &3.70 &3.99& 5.78 & 76.98&75.62 &71.36  &3.66 &5.96 & 63.21 & 77.06 &75.07&71.21 &3.15&4.25&6.25 &77.27 &74.83 & 71.22 & 3.47&4.90&\underline{6.61} &\underline{78.87} &73.65 & 70.13 &9.57  & 74.43 \\     
          &    RLR  & 80.64 &85.17& 87.65 &17.63 &13.20& 10.18 &11.54 &15.61&  30.51 & 29.98 &22.58& 14.04 &8.54&13.00&14.15 & 30.60&24.94 &16.49 &76.35 & 88.80& 95.34& 22.30 &9.35& 2.09& 50.60 & 17.78 \\
          & RFA   & 73.73 &100& 100  & 7.51&0.00 &0.00 &84.22 & 100& 100 & 6.27 &0.00&0.00 &93.60&100& 100& 1.39& 0.00& 0.00 & 78.41&100 &100   & 5.09 &0.00& 0.00 &94.16  & 1.68 \\
          &     MMetric & 93.70 &93.49& 99.63 &5.22 &4.84& 0.32 & 53.90& 23.61& 67.23 & 35.79 &53.99&27.59 &99.37&99.96&99.87 & 0.36&0.04 &0.13  &91.73 &89.74& 98.42 & 7.52 &9.53&0.57  &  84.22 &12.15 \\
&     RoseAgg & 97.31&99.36&  100& 0.07&0.01& 0.00 & 92.54& 94.27& 98.65 & 6.51 & 4.36&0.98  & 72.64& 78.40& 85.83 & 25.71& 20.96 &  13.25 &  84.69& 90.10& 91.63  & 14.02  &8.53 &  8.02 & 90.45  &  8.53\\
          &  AlignIns &\textbf{2.88} &\underline{3.20}& \underline{5.71} & \textbf{79.99} &\textbf{80.13} & \underline{72.19} & \underline{3.17}& \textbf{3.07}& \underline{6.97} & \textbf{79.67} &\textbf{79.74}&\underline{75.64} &\underline{3.11}&\textbf{2.99}&\underline{6.02} &\textbf{79.94} &\textbf{79.87} &\underline{76.57} & \underline{3.32}& \underline{4.12}& 8.69& 77.64 &\underline{76.17}& \underline{70.49}&\underline{4.43}  &\underline{77.33}\\
                              &  EndPCA &2.73 &4.21& 6.26 & 78.96 &76.20 & 73.54  & 3.76& 3.31& 5.43 & 77.78 &77.16&75.90 &4.64&5.31&7.19&76.82 &75.03 &74.74  &3.85& 5.47& 8.64& 78.02 &76.86& 75.15&5.06 & 76.34 \\
          & \cellcolor[rgb]{ .816,  .808,  .808}FedDAB & \cellcolor[rgb]{ .816,  .808,  .808}\underline{3.22} &\cellcolor[rgb]{ .816,  .808,  .808}\textbf{2.97}& \cellcolor[rgb]{ .816,  .808,  .808}\textbf{3.02}&\cellcolor[rgb]{ .816,  .808,  .808}\underline{78.91} &\cellcolor[rgb]{ .816,  .808,  .808}\underline{78.93} &\cellcolor[rgb]{ .816,  .808,  .808}\textbf{78.26} &\cellcolor[rgb]{ .816,  .808,  .808}\textbf{2.99} &\cellcolor[rgb]{ .816,  .808,  .808}\underline{3.09} & \cellcolor[rgb]{ .816,  .808,  .808}\textbf{3.87}& \cellcolor[rgb]{ .816,  .808,  .808}\underline{79.59} &\cellcolor[rgb]{ .816,  .808,  .808}\underline{78.83}&\cellcolor[rgb]{ .816,  .808,  .808}\textbf{77.08} &\cellcolor[rgb]{ .816,  .808,  .808}\textbf{2.82}&\cellcolor[rgb]{ .816,  .808,  .808}\underline{3.02}&\cellcolor[rgb]{ .816,  .808,  .808}\textbf{3.73} & \cellcolor[rgb]{ .816,  .808,  .808}\underline{79.67}&\cellcolor[rgb]{ .816,  .808,  .808}\underline{78.97} & \cellcolor[rgb]{ .816,  .808,  .808}\textbf{78.08} & \cellcolor[rgb]{ .816,  .808,  .808}\textbf{3.10}& \cellcolor[rgb]{ .816,  .808,  .808}\textbf{2.93}& \cellcolor[rgb]{ .816,  .808,  .808}\textbf{3.66} & \cellcolor[rgb]{ .816,  .808,  .808}\textbf{79.07} &\cellcolor[rgb]{ .816,  .808,  .808}\textbf{79.19}&\cellcolor[rgb]{ .816,  .808,  .808}\textbf{77.90} &\cellcolor[rgb]{ .816,  .808,  .808}\textbf{3.20}& \cellcolor[rgb]{ .816,  .808,  .808}\textbf{78.71} \\\midrule 
        \multirow{8}[4]{*}{\rotatebox{90}{\makecell*[c]{CIFAR100 \\ (VGG)}}} & FedAvg & 81.94 &96.64& 98.14 & 6.51&2.47 &0.54  &64.63 & 69.23 &71.32 &16.90  &15.27 &14.80 &62.75&72.73& 75.61&16.83 &10.55 &8.18 & 79.05& 96.76& 98.15&7.56  &2.29& 0.37  &80.57 & 8.52 \\ \cmidrule{2-28}  
                  & MKrum & 66.48 &94.47& 98.74 & 14.64 &3.48 & 0.27  &54.66& 93.65& 97.40&21.33  &4.29&0.57 &86.46&94.47&96.16 &4.79 &3.48 & 2.90 & 82.62&95.00& 98.37 & 9.66 &4.07& 1.80 & 88.20 & 5.93 \\
          & Foolsgold  &3.75 &\underline{1.96}& \underline{4.25} & \underline{45.65}&43.41 &\underline{40.62}  &5.25 &2.46 & 8.15 & 45.03 &43.87&35.97 &\underline{0.72}&\underline{0.94}&\underline{1.08} &\underline{46.11} &43.61 & \underline{42.90} & 35.00&\underline{1.15}&10.83 &9.53 &44.61 & 30.66 &6.29  & 39.33 \\      
          &    RLR  & 63.37 &100& 100 &0.49 &0.00& 0.00 &11.54 &32.67 & 40.74 & 32.98 &11.58& 4.04 &8.54&60.50&81.10 & 30.60&14.94 &6.49 &21.95 & 29.63& 36.00& 22.61 &10.35& 5.09& 48.83 & 11.59 \\
          & RFA   & 73.73 &85.50& 87.64 & 7.51&5.36 &5.17 &67.31 & 84.22& 90.43 & 6.27 &2.61&2.29 &75.31&84.42& 87.06& 1.63& 1.27& 0.95 & 52.81&89.92 &93.15   & 5.09 &0.62& 0.31 &80.95  & 3.25 \\
          &     MMetric & 92.14 &99.98& 100 &5.22 &0.00& 0.00 & 99.63& 99.83& 100 & 0.32 &0.25&0.00 &24.53&43.80&46.80& 16.44&9.63 &1.54 &64.99 & 70.64& 72.97 & 10.33&5.47&3.99  &  76.27 &4.43 \\
&     RoseAgg & 95.48&99.64&100  &0.67 &0.30& 0.00 & 92.61& 95.37&99.70  & 1.32 &0.77 &0.37  &74.66 & 86.16 & 99.54 &2.74 & 1.60 & 0.13  & 79.73 &85.80 & 96.74  & 2.87  &1.65 & 0.96  &  92.11 & 1.12 \\
          &  AlignIns &\underline{1.45} &2.98& 100 & 45.45 &\underline{44.03} & 0.00  & \underline{1.35}& \underline{1.40}& \underline{1.87} & \underline{47.44}&\underline{44.97}&\underline{43.20} &1.13&1.28&1.59 &45.26 &\underline{44.29}&42.61 & \underline{0.81}& 1.25& \underline{1.41}& \underline{45.80} &\underline{44.66}& \underline{42.30}&9.71 & 40.83 \\
                              &  EndPCA &1.74 &3.15& 9.53 & 45.10 &42.48 & 39.29  & 1.95& 1.82& 7.27 & 45.56 &44.70&41.19 &0.82&2.86&8.59&46.42 &43.78 &38.50 &2.77& 3.69& 6.08& 44.01 &43.85& 40.67&\underline{4.18}  & \underline{42.96} \\
          & \cellcolor[rgb]{ .816,  .808,  .808}FedDAB & \cellcolor[rgb]{ .816,  .808,  .808}\textbf{0.67} &\cellcolor[rgb]{ .816,  .808,  .808}\textbf{0.26}& \cellcolor[rgb]{ .816,  .808,  .808}\textbf{0.42} &\cellcolor[rgb]{ .816,  .808,  .808}\textbf{47.69} &\cellcolor[rgb]{ .816,  .808,  .808}\textbf{46.33} &\cellcolor[rgb]{ .816,  .808,  .808}\textbf{44.61}  &\cellcolor[rgb]{ .816,  .808,  .808}\textbf{0.71} &\cellcolor[rgb]{ .816,  .808,  .808}\textbf{0.32} & \cellcolor[rgb]{ .816,  .808,  .808}\textbf{0.41} & \cellcolor[rgb]{ .816,  .808,  .808}\textbf{47.91} &\cellcolor[rgb]{ .816,  .808,  .808}\textbf{46.24}&\cellcolor[rgb]{ .816,  .808,  .808}\textbf{44.64} &\cellcolor[rgb]{ .816,  .808,  .808}\textbf{0.70}&\cellcolor[rgb]{ .816,  .808,  .808}\textbf{0.29}&\cellcolor[rgb]{ .816,  .808,  .808}\textbf{0.43} & \cellcolor[rgb]{ .816,  .808,  .808}\textbf{47.91}&\cellcolor[rgb]{ .816,  .808,  .808}\textbf{46.38} & \cellcolor[rgb]{ .816,  .808,  .808}\textbf{44.63} & \cellcolor[rgb]{ .816,  .808,  .808}\textbf{0.73}& \cellcolor[rgb]{ .816,  .808,  .808}\textbf{0.23}& \cellcolor[rgb]{ .816,  .808,  .808}\textbf{0.42}& \cellcolor[rgb]{ .816,  .808,  .808}\textbf{48.15} &\cellcolor[rgb]{ .816,  .808,  .808}\textbf{46.33}&\cellcolor[rgb]{ .816,  .808,  .808}\textbf{44.61}  & \cellcolor[rgb]{ .816,  .808,  .808}\textbf{0.46} & \cellcolor[rgb]{ .816,  .808,  .808}\textbf{46.28} \\\midrule\midrule
    \end{tabular}%
    }
  \label{tab:addlabel}%
\end{table*}%

\subsubsection{Comparison with Existing Methods}
Table \ref{tab:addlabel} reports the performance of FedDAB and existing defense methods against backdoor attacks under the $\operatorname{Dir}(1.0)$ setting. 
The ASR and RR values are recorded at the round with the highest Top-1 test accuracy.
Overall, FedDAB demonstrates superior defense performance under  four backdoor attacks.
Specifically, on FMNIST, Foolsgold achieves the highest RR (67.36\%), but its ASR remains higher than that of FedDAB. 
This is  because Foolsgold cannot  filter out malicious updates and can only reduce their weights in global aggregation. 
Consequently, malicious updates are  incorporated into the aggregation process, resulting in a relatively high ASR. 
In addition, RoseAgg exhibits the highest ASR. 
This is because RoseAgg evaluates local updates based on their directions, making its assessment susceptible to deviations among benign update directions induced by Non-IID data, preventing it from assigning  low aggregation weights to malicious updates.
The defense effectiveness of the other methods is also impaired by statistical heterogeneity, making them less robust than FedDAB against these backdoor attacks.
More specifically, on CIFAR10, FedDAB reduces the ASR by 1.23 \%  and improves the RR by 1.38 \% over the second-best method. Similarly, on CIFAR100, FedDAB outperforms all existing methods in terms of both ASR and RR.
The performance advantages of FedDAB can be attributed to two key designs. $(\textit{i)}$ The model-contrastive term improves the consistency of benign updates in both direction and magnitude, enabling the alignment checking to distinguish benign updates from malicious ones more accurately. $(\textit{ii)}$ FedDAB employs both overall-direction checking and parameter-level checking to evaluate local updates from complementary perspectives, making it more difficult for malicious updates to evade detection.

\begin{table}[t]
  \centering
  \caption{The  ASR (\%)  results of FedDAB on     various  Non-IID settings of FMNIST, CIFAR10, and CIFAR100. }
  \label{tab:main_results}
\resizebox{\columnwidth}{!}{
      \renewcommand{\arraystretch}{0.6}
\tabcolsep=0.004cm
    \begin{tabular}{c|c|ccc|ccc|ccc|ccc|c}
\midrule\midrule
    \multirow{2}[4]{*}{\textbf{\makecell*[c]{Dataset \\ (Model)}}} & \multirow{2}[4]{*}{$\operatorname{Dir}(\Psi)$}  & \multicolumn{3}{c|}{\textbf{BadNet}}  &     \multicolumn{3}{c|}{\textbf{DBA}}  &    \multicolumn{3}{c|}{\textbf{Neurotoxin}}  &     \multicolumn{3}{c|}{\textbf{PGD}}& \multirow{2}{*}{\textbf{\makecell{Avg. \\ ASR$\downarrow$}}} \\
\cmidrule(r){3-5} \cmidrule(r){6-8} \cmidrule(r){9-11} \cmidrule(r){12-14} &             & $v$=0.30 & $v$=0.40& $v$=0.50 & $v$=0.30 & $v$=0.40& $v$=0.50 & $v$=0.30 & $v$=0.40& $v$=0.50 & $v$=0.30 & $v$=0.40& $v$=0.50& \\\midrule
        \multirow{3}[4]{*}{\rotatebox{360}{\makecell*[c]{FMNIST \\ (ResNet9)}}} & $\Psi = 0.8$ & 3.91 &6.97&   3.79 & 4.77&4.23 &3.81 &3.59 & 3.57 &4.41 &2.72  &4.03 &4.22& 4.17\\ \cmidrule{2-15}        
          &    $\Psi = 0.5$  & 8.70 &4.63& 4.52 &3.68 &3.86& 3.12 &6.21 &4.79 & 4.67 & 5.10 &4.46& 3.96 &4.81
         \\ \cmidrule{2-15}   
                   &    $\Psi = 0.3$  &10.14 &3.08& 4.49 &8.15 &7.24& 3.16 &4.03 &2.14 & 3.24 & 4.64 &1.76& 8.03 
 &5.01
         \\ \midrule  
            \multirow{3}[4]{*}{\rotatebox{360}{\makecell*[c]{CIFAR10 \\ (ResNet9)}}} & $\Psi = 0.8$ & 3.82 &3.07& 3.94 & 3.91&3.77 &3.44  &3.91 &4.52 &3.59 &3.67 &3.09 &3.50 &3.69\\ \cmidrule{2-15}        
          &    $\Psi = 0.5$  & 3.98 &4.17& 6.23 &4.10 &4.01& 6.30 &4.48 &3.89 & 6.14 & 4.12 &4.24& 6.27 
 &4.83
         \\ \cmidrule{2-15}   
                   &    $\Psi = 0.3$  & 2.56 &3.16& 6.06 &2.51 &3.17& 5.29 &3.90 &4.04 & 5.74 & 2.59 &3.81& 6.60 
 &4.12
         \\ \midrule
            \multirow{3}[4]{*}{\rotatebox{360}{\makecell*[c]{CIFAR100 \\ (VGG)}}} & $\Psi = 0.8$ & 0.58 &1.18& 0.84 &0.98&1.17 &0.92  &0.42 & 1.40 &0.96 &0.53  &1.30 &0.82  &0.93
         \\ \cmidrule{2-15}         
          &    $\Psi = 0.5$  & 1.05 &0.76& 0.69 &0.89 &0.82& 0.90 &0.98&0.82 &0.85 & 0.97 &0.75& 0.93 
 &0.87
         \\ \cmidrule{2-15}    
                   &    $\Psi = 0.3$  & 0.15 &0.49& 0.87 &0.19 &0.90& 0.86 &0.14 &0.56 & 0.76  & 0.04 &0.49&0.94 
 &0.53
         \\ 
    \midrule\midrule
    \end{tabular}%
    }
  \label{tab:addlabel123}%
\end{table}%

\subsubsection{Effectiveness of FedDAB in Varying  Non-IID Settings}
To evaluate the performance of FedDAB under different data distributions, we conduct experiments under three Non-IID settings: $\operatorname{Dir}(0.8)$, $\operatorname{Dir}(0.5)$, and $\operatorname{Dir}(0.3)$.
Table \ref{tab:addlabel123} reports the ASR achieved by FedDAB.
Across the four backdoor attacks, the ASR exhibits some fluctuations as statistical heterogeneity increases.
Nevertheless, FedDAB  maintains its effectiveness  across different Non-IID settings.
Furthermore, FedDAB achieves lower ASR on CIFAR-10/100 than on FMNIST.
This can be attributed to the more complex sample characteristics of CIFAR-10/100,  which make it harder for malicious nodes to manipulate the global model toward their backdoor objectives.
Moreover, on CIFAR-10/100, the ASR changes  marginally   when the proportion of malicious nodes increases from 30\% to 50\%,  indicating that FedDAB is robust to variations in the proportion of malicious nodes.
This conclusion is further supported by the convergence curves in Fig. \ref{1121011234}.

To evaluate the performance of FedDAB under extreme Non-IID settings, we  conduct  experiments under the $\operatorname{Dir}(0.1)$ setting,  using the relatively moderate $\operatorname{Dir}(0.5)$ setting as a reference.
The results are reported in Table \ref{21727124}. 
We observe that, despite the  higher degree of statistical heterogeneity under $\operatorname{Dir}(0.1)$, the ASR of FedDAB remains comparable to that under $\operatorname{Dir}(0.5)$, without any noticeable increase.
 These results demonstrate that FedDAB can effectively mitigate backdoor attacks even under extreme Non-IID conditions.

\begin{table}[t]
  \centering
  \caption{Performance of FedDAB  with  extreme Non-IID Settings.}
       \renewcommand{\arraystretch}{0.4}
\resizebox{\columnwidth}{!}{
    \begin{tabular}{c|c|cc|cc|cc|cc}
\midrule\midrule
\multirow{3}{*}{\textbf{Dataset}} &\multirow{3}{*}{$v$} &\multicolumn{4}{c|}{BadNet}&\multicolumn{4}{c}{DBA } \\
&  
& \multicolumn{2}{c|}{$\operatorname{Dir(0.5)}$} 
& \multicolumn{2}{c|}{$\operatorname{Dir(0.1)}$} 
& \multicolumn{2}{c|}{$\operatorname{Dir(0.5)}$} 
& \multicolumn{2}{c}{$\operatorname{Dir(0.1)}$} \\
&  & ASR$\downarrow$ & RR$\uparrow$   & ASR$\downarrow$ & RR$\uparrow$  & ASR$\downarrow$   & RR$\uparrow$  & ASR$\downarrow$ & RR$\uparrow$\\
    \midrule
\multirow{3}{*}{CIFAR10}  &  20\%   & 3.64 & 74.15 & 2.17 & 70.71& 2.65 &75.84 &3.58& 68.19 \\
   & 30\%  & 4.58 & 72.99 & 4.01 &  68.19 & 3.32 & 73.11 &5.64& 67.42 \\
   &  40\% & 4.23&  71.01 & 5.84&  68.92& 2.83& 70.66& 3.45& 65.40\\
         \midrule
\multirow{3}{*}{CIFAR100}  & 20\%   &0.43 & 42.83 & 0.19 & 35.88& 0.54 & 43.46 &0.59& 38.48 \\
   & 30\%  & 0.15 &41.91 & 0.86 &33.42 & 0.58 & 43.87 & 0.97&36.68 \\
   & 40\%  & 0.29& 40.62 & 1.10&  32.40& 0.71 & 41.27& 0.93& 33.45\\
\midrule\midrule
    \end{tabular}%
    }
  \label{21727124}%
\end{table}%

\begin{table}[t]
\centering
\caption{Impact of each component in FedDAB under CIFAR100.}
       \renewcommand{\arraystretch}{0.4}
\resizebox{\columnwidth}{!}{
\tabcolsep=0.004cm
    \centering
\begin{tabular}{c|cc|cc|cc|cc|c}
\midrule\midrule
  \multirow{3}{*}[-1ex]{\textbf{Configuration}}  &\multicolumn{8}{c|}{\textbf{Attack Success Rate} (\%) $\downarrow$}  &  \multirow{3}{*}[0ex]{\makecell{Avg. \\ ASR$\downarrow$}} \\
&\multicolumn{2}{c|}{BadNet}    &\multicolumn{2}{c|}{DBA}    &\multicolumn{2}{c|}{Neurotoxin}   &\multicolumn{2}{c|}{PGD}  &\\
&$\operatorname{Dir}(1.0)$&$\operatorname{Dir}(0.5)$&$\operatorname{Dir}(1.0)$ &$\operatorname{Dir}(0.5)$&$\operatorname{Dir}(1.0)$&$\operatorname{Dir}(0.5)$ &$\operatorname{Dir}(1.0)$&$\operatorname{Dir}(0.5)$&\\
\midrule
$\operatorname{DSS}$&2.24&20.95 &2.61& 31.60&2.70 & 25.78&2.48& 33.35&15.21 \\
$\quad \quad\quad \raisebox{0.4em}{\rule{0.4pt}{0.5em}}\raisebox{0.3em}{\rule{0.8em}{0.4pt}}+ \operatorname{MCT}$  &1.95&12.30 &1.77&21.84 &2.09& 17.63&2.15& 13.77& 9.18\\\midrule
$\operatorname{SAS}(\text{Top}$-$30\%)$&1.65& 12.65 &1.48 & 12.31&1.26&11.97 &1.01& 12.92&6.90\\
$\quad \quad\quad\raisebox{0.4em}{\rule{0.4pt}{0.5em}}\raisebox{0.3em}{\rule{0.8em}{0.4pt}}+ \operatorname{MCT}$&1.21& 9.18&1.83 & 8.90 &1.55&10.24 &1.58&12.37 & 5.85 \\\midrule
$\operatorname{DSS}+\operatorname{SAS}(\text{Top}$-$100\%)$&1.77 &5.70 &1.63& 6.40 &2.04& 5.35&1.75&8.50 & 4.14 \\
$\quad \quad\quad\raisebox{0.4em}{\rule{0.4pt}{0.5em}}\raisebox{0.3em}{\rule{0.8em}{0.4pt}}+ \operatorname{MCT}$&0.82 &2.18 &0.96& 1.91&0.89& 2.24&0.84&2.37 & 1.52\\\midrule
$\operatorname{DSS}+\operatorname{SAS}(\text{Top}$-$50\%)$&1.40&1.44 &1.38&1.29 &1.07& 1.13&0.92& 1.10& 1.21\\
$\quad \quad\quad\raisebox{0.4em}{\rule{0.4pt}{0.5em}}\raisebox{0.3em}{\rule{0.8em}{0.4pt}}+ \operatorname{MCT}$&0.83&1.08 &0.65& 0.95 &0.72&0.84 &0.68&0.98 & 0.84\\\midrule
FedDAB $(H = 0)$&1.24& 1.30 & 1.08&1.14 &1.15&1.06 &0.84&1.25 &1.13 \\
FedDAB $(H = 3)$&0.67&1.05  & 0.71& 0.89&0.70&0.98 &0.73&0.94 & 0.83\\\midrule
\midrule
\end{tabular}}
\label{3159191351}
\end{table}

\subsubsection{Ablation Study}

FedDAB incorporates a model-contrastive term ($\operatorname{MCT}$) into the local training objective and computes the $\operatorname{DSS}$ and $\operatorname{SAS}$ of each node to identify and filter malicious updates.
To investigate the contribution of each component, we conduct comprehensive ablation studies on CIFAR100 with 30\% malicious nodes.
The results are reported in Table~\ref{3159191351}, where $\operatorname{SAS}$ (Top-$r$\%) indicates that  the top $r\%$ of model parameters  are used for parameter-level checking.
(\textit{i}) Component ablation.
We observe that under the $\operatorname{Dir(1.0)}$ setting, either $\operatorname{DSS}$  or $\operatorname{SAS}$ alone achieves ASR  comparable to that of FedDAB.
This is because  the relatively balanced data distribution induced by $\operatorname{Dir}(1.0)$ leads to  consistent benign updates, making a single check sufficient to provide strong robustness.
However, under the $\operatorname{Dir(0.5)}$ setting, benign updates exhibit substantial deviation, which significantly reduces the effectiveness of relying on either $\operatorname{DSS}$ or $\operatorname{SAS}$ alone.
When the two components are combined, $\operatorname{DSS}+\operatorname{SAS}$ (\text{Top-}$100\%$) reduces the average ASR to 4.14\%, demonstrating their complementary strengths.
In addition, incorporating $\operatorname{MCT}$ consistently reduces the ASR across different configurations.
This improvement can be attributed to the model-contrastive term, which enhances the consistency of benign updates, improving the robustness of alignment checking process.
(\textit{ii}) Important parameter ablation.
To examine the effect of selecting important parameters, we evaluate $\operatorname{DSS}+\operatorname{SAS}$ using either all parameters (Top-100\%) or only  important parameters (Top-50\%).
When only the top-50\% of parameters are used, the ASR is reduced compared with the top-100\% setting, regardless of whether $\operatorname{MCT}$ is applied.
These results indicate that restricting $\operatorname{SAS}$ to the important parameters of local updates can improve the accuracy of filtering malicious updates.
(\textit{iii}) Historical information  ablation.
Since   benign nodes generally exhibit consistent and honest behavior across training rounds, FedDAB incorporates their historical information into parameter-level checking.
Specifically, we set the length of the global sign buffer $\mathcal{S}_g$ and the local sign buffer $\mathcal{S}_l$ to $H$, , where $H=0$ indicates that no historical information is used.
 Under both the $\operatorname{Dir}(0.5)$ and  $\operatorname{Dir}(1.0)$ settings, FedDAB ($H$=$3$) achieves a lower   ASR  than  FedDAB ($H$=$0$).
This result indicates that leveraging historical behaviors improves FedDAB's robustness against backdoor attacks.

\begin{figure}[t]
  \centering
  \subfloat[Neurotoxin  Attacks.]
  {
      \label{2063141}  \includegraphics[width=0.3\linewidth]{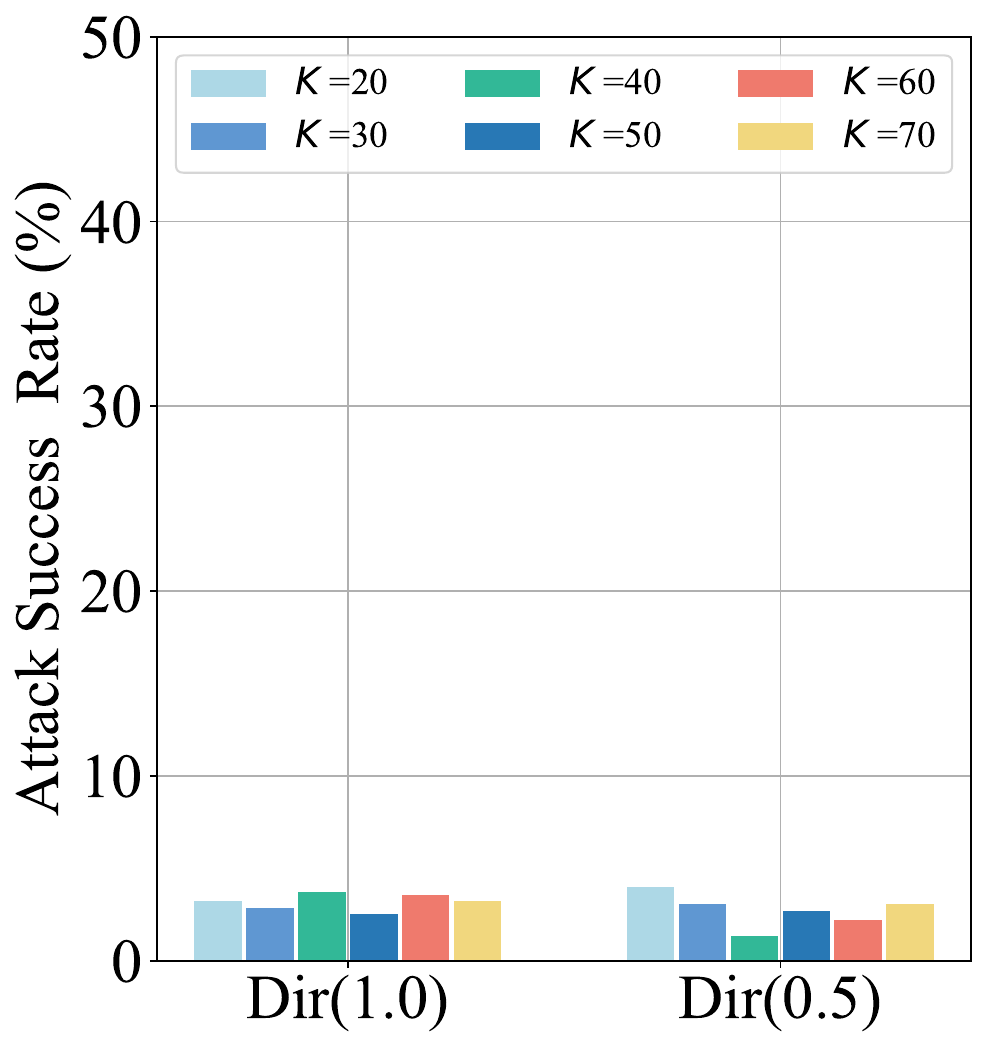}
  }
  \subfloat[Neurotoxin  Attacks.]
  {
      \label{20631412}  \includegraphics[width=0.3\linewidth]{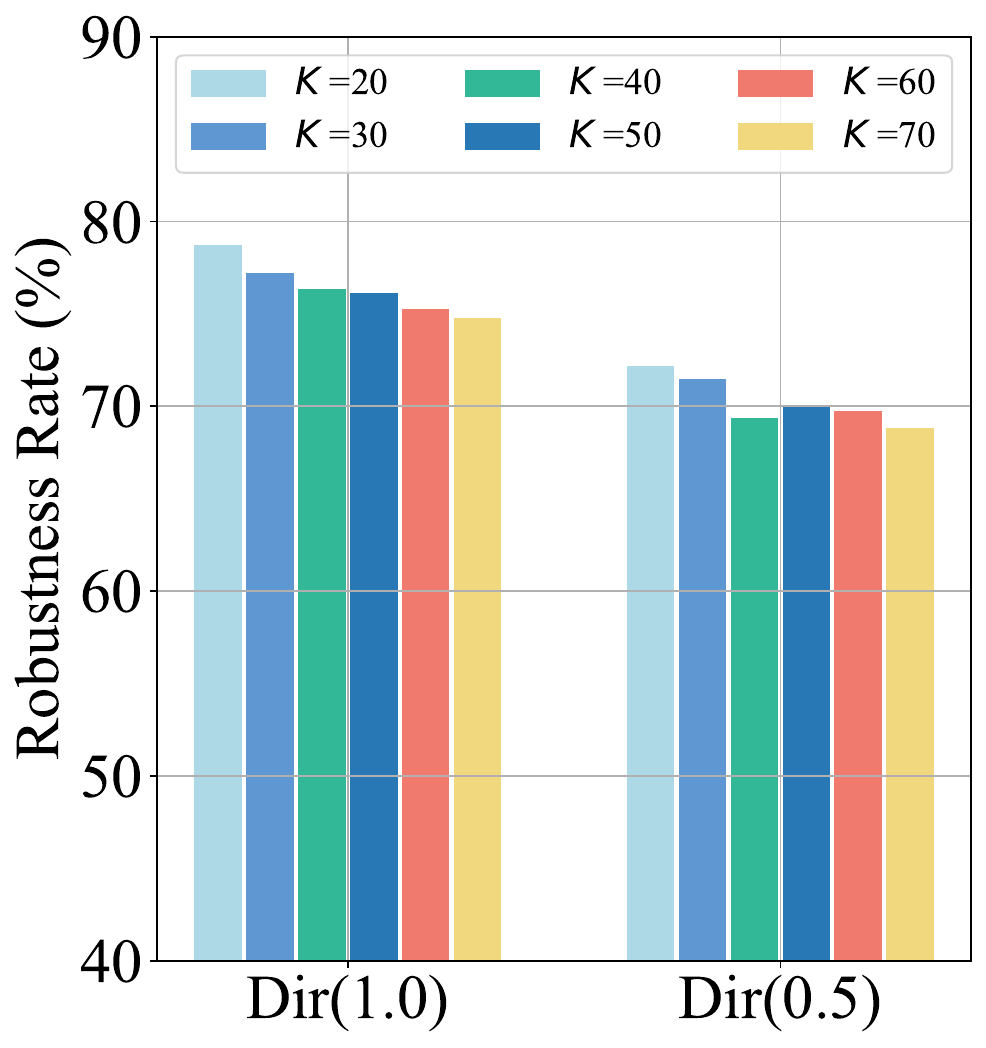}
  }
    \subfloat[Neurotoxin  Attacks.]
  {
      \label{20631413}  \includegraphics[width=0.3\linewidth]{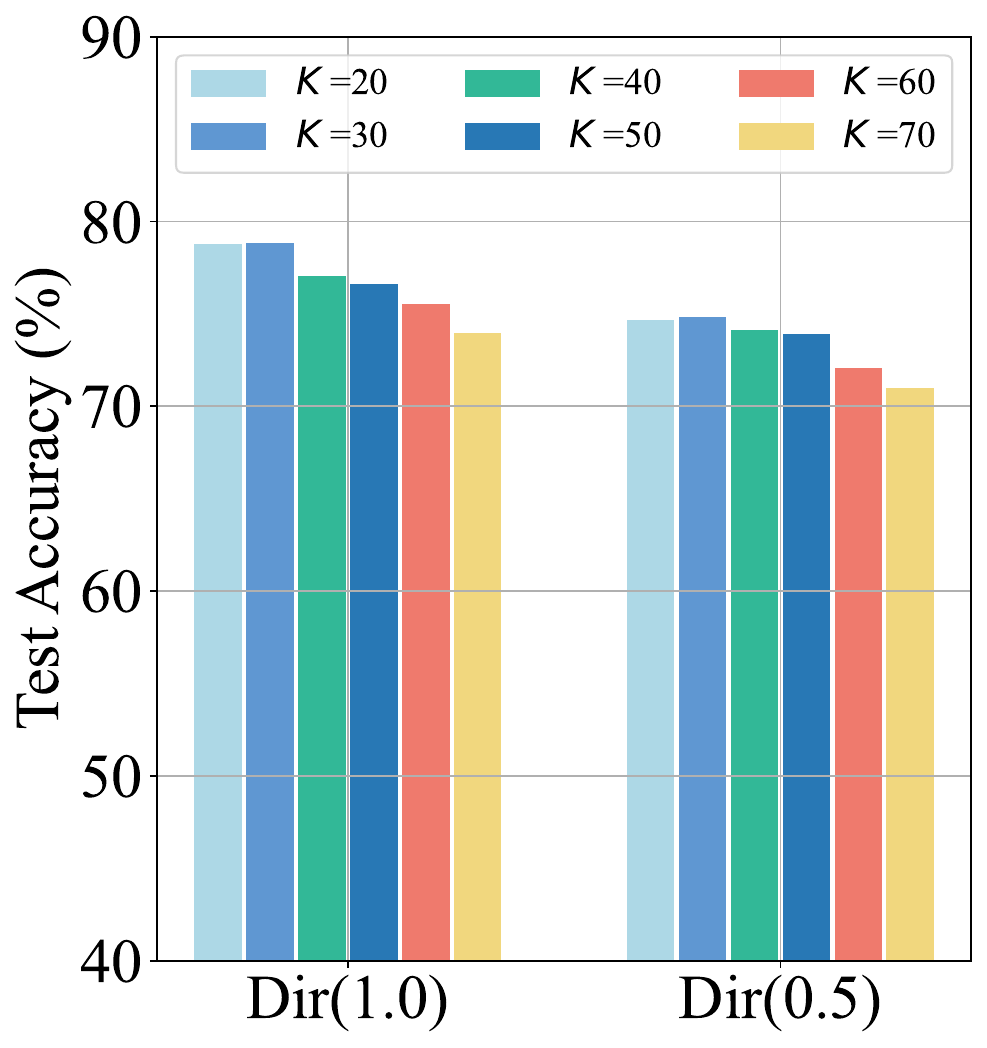}
  }
  
    \subfloat[PGD Attacks.]
  {
      \label{2063141}  \includegraphics[width=0.3\linewidth]{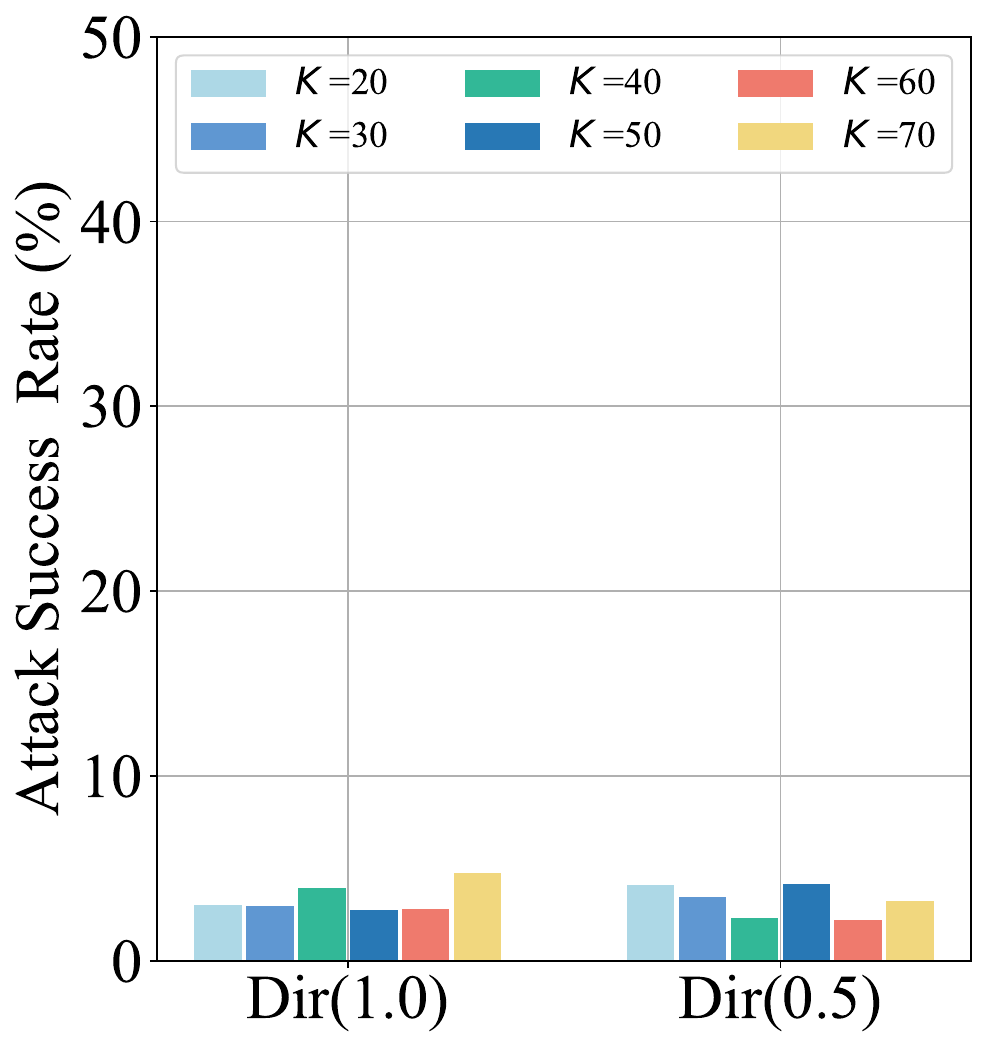}
  }
  \subfloat[PGD Attacks.]
  {
      \label{20631412}  \includegraphics[width=0.3\linewidth]{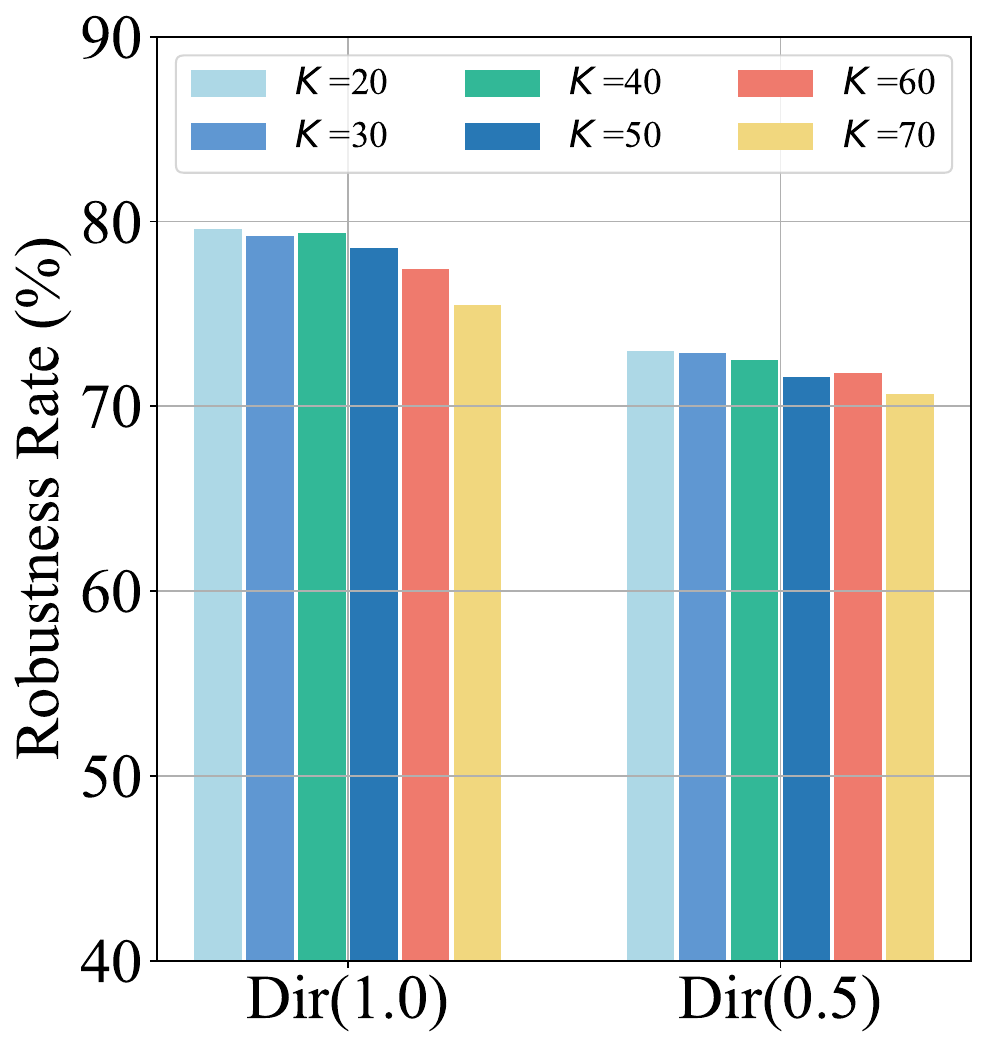}
  }
    \subfloat[PGD Attacks.]
  {
      \label{20631413}  \includegraphics[width=0.3\linewidth]{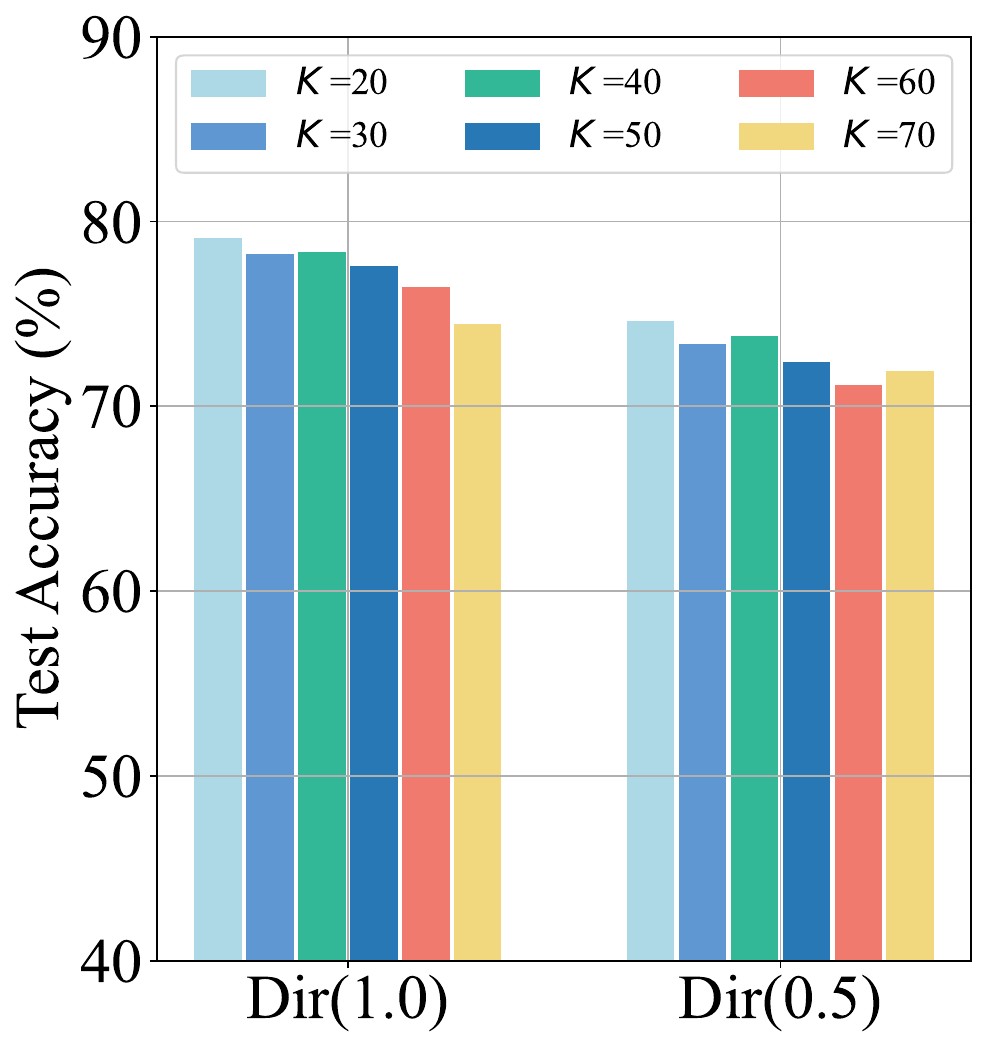}
  }
\caption{Test performance of FedDAB on CIFAR10 with different numbers of nodes under Neurotoxin and PGD attacks.}
\label{2063141300}
\end{figure}

\subsubsection{FedDAB with Varying Numbers of Nodes} \label{561449}
In real-world federated learning scenarios, the number of participating nodes is often not fixed.
Therefore, it is essential to evaluate the scalability of FedDAB under different system scales.
To this end, we evaluate FedDAB's ASR, RR, and TA metrics on CIFAR10    with 20, 30, 40, 50, 60, and 70 nodes, where  30\% of  the nodes are malicious.
Notably, as the number of nodes increases, the default number of rounds $T$ becomes insufficient to ensure model convergence at larger scales.
To ensure    model convergence across all system scales, we set the number of  rounds $T$ to 300.
The experimental results are shown  in Fig. \ref{2063141300}.
Under both the $\operatorname{Dir}(1.0)$ and $\operatorname{Dir}(0.5)$ settings, the ASR does not increase as the number of nodes grows.
The results indicate  that FedDAB effectively prevents malicious nodes from achieving their attack objectives, demonstrating its strong scalability across different system scales.
In addition, the RR and TA decrease slightly with the number of participating nodes.
This is primarily because  a larger number of nodes leads to more dispersed data distributions, which negatively impacts  the global model's performance.
Notably, this phenomenon is   common in large-scale federated learning systems.
 Overall, FedDAB maintains robust defensive capability across FL systems of different scales.

\subsubsection{Stability of  FedDAB with Varying Backdoor Attacks}
\begin{figure}[!t]
  \centering
  \subfloat[$\operatorname{Dir}(1.0)$.]
  {
      \label{206314113}  \includegraphics[width=0.45\linewidth]{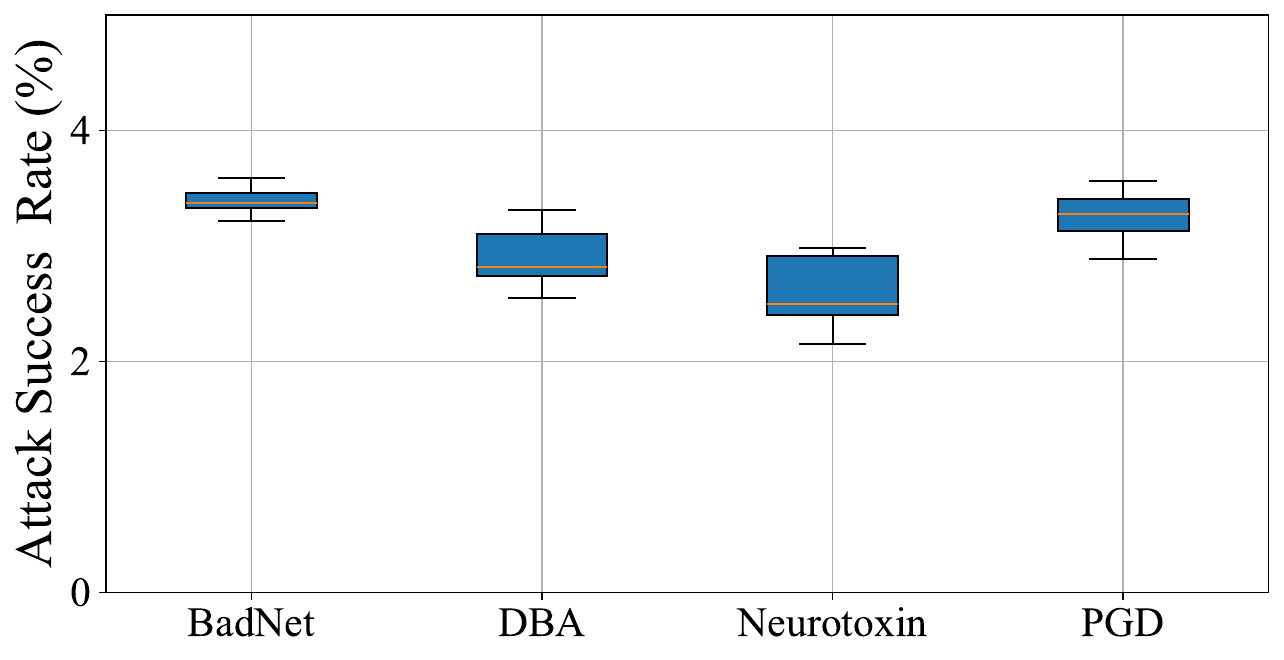}
  }
  \subfloat[$\operatorname{Dir}(0.5)$.]
  {
      \label{20631412445}  \includegraphics[width=0.45\linewidth]{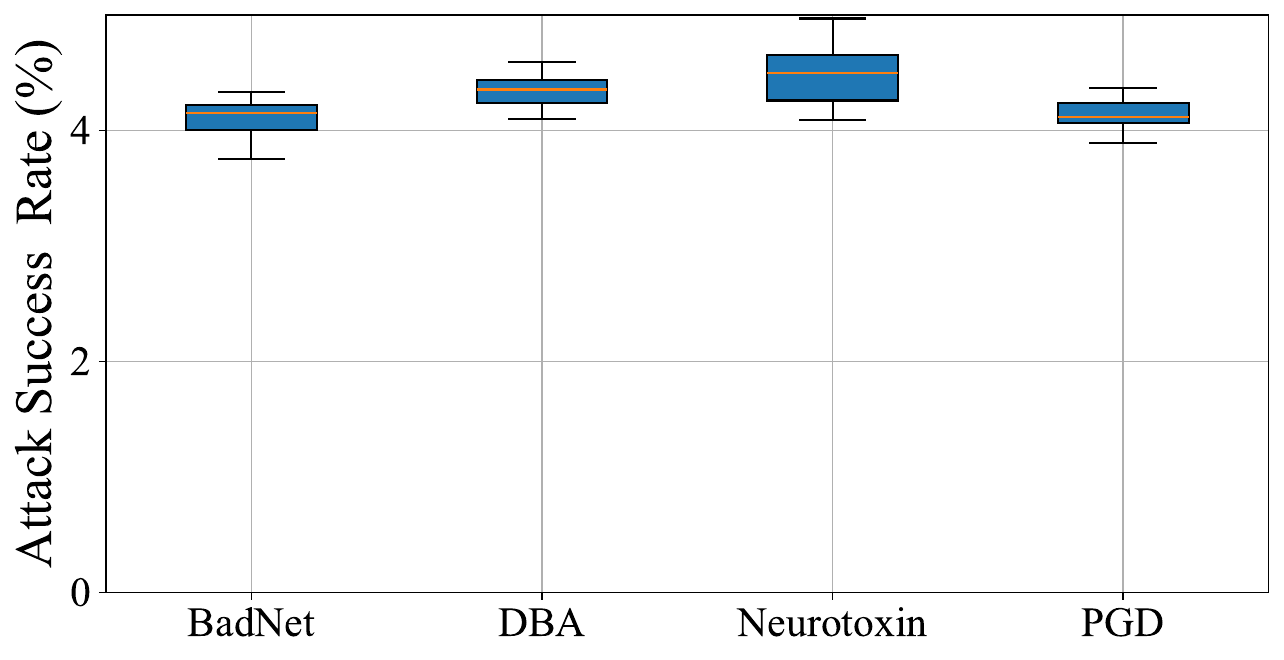}
  }
  
    \subfloat[$\operatorname{Dir}(1.0)$.]
  {
      \label{206314113}  \includegraphics[width=0.45\linewidth]{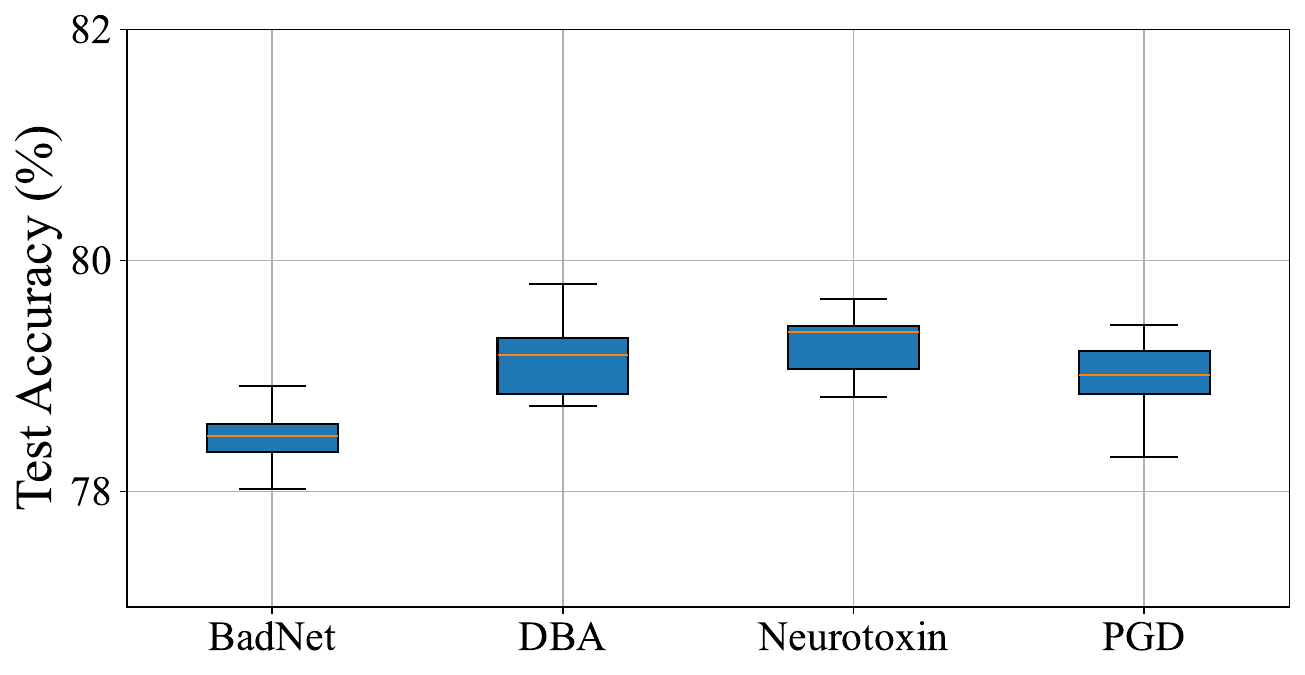}
  }
  \subfloat[$\operatorname{Dir}(0.5)$.]
  {
      \label{20631412445}  \includegraphics[width=0.45\linewidth]{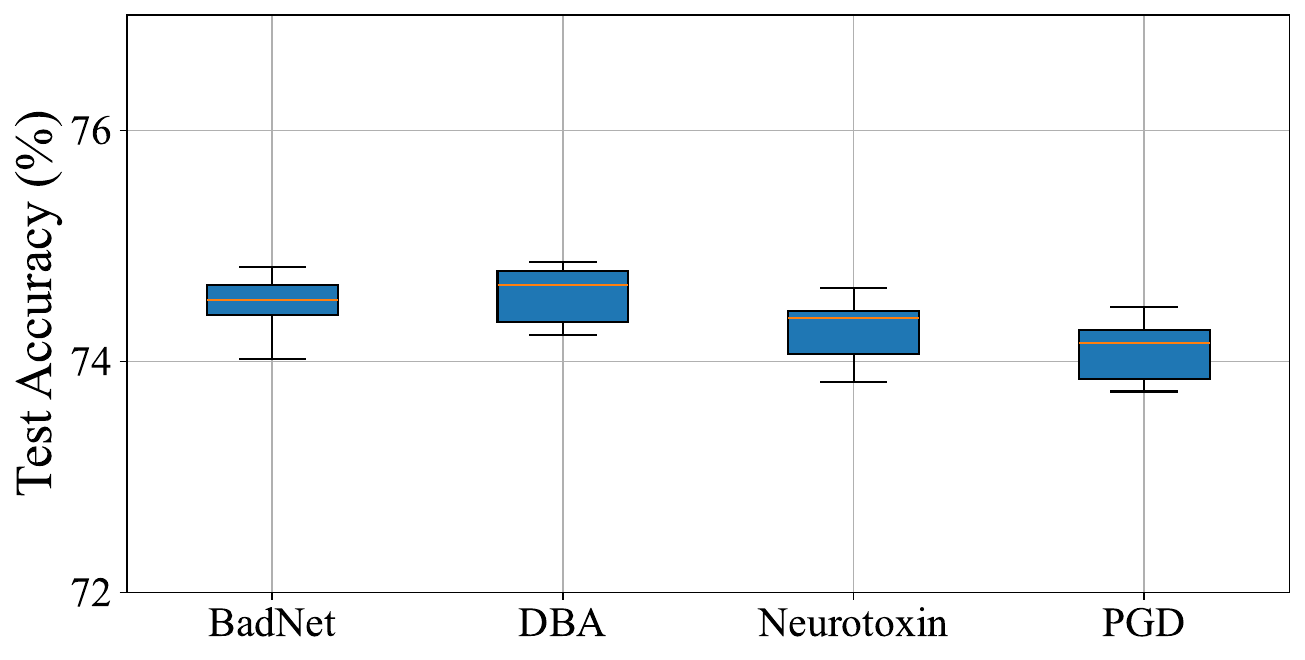}
  }
\caption{Box plots of the  stability of FedDAB against various   backdoor attacks under CIFAR10 dataset with $\operatorname{Dir}(1.0)$ and $\operatorname{Dir}(0.5)$ settings.}
\label{2063141300123}
\end{figure}

Ensuring the stability of results across repeated experiments is particularly important in federated learning systems.
 To this end, we conduct ten repeated experiments on the CIFAR10  under both $\operatorname{Dir}(1.0)$ and $\operatorname{Dir}(0.5)$ settings to evaluate the ASR and TA of FedDAB against different backdoor attacks.
 All hyper-parameters are kept at their default values.
 The  results are presented in Fig. \ref{2063141300123}.
 FedDAB maintains high stability across various backdoor attacks.
 The ASR fluctuations are consistently controlled within 0.5\%, and the TA fluctuations are consistently controlled within 1\% over repeated runs.
These  variations primarily result  from the inherent randomness of gradient descent algorithms, leading to  differences in the learned global model. 
Thus, the results demonstrate that FedDAB maintains stable defense performance across multiple independent runs.

\subsubsection{Convergence of  FedDAB under Different Malicious  Proportions}
\begin{figure}[!t]
  \centering
  \subfloat[BadNet  Attacks.]
  {
      \label{1121010}  \includegraphics[width=0.45\linewidth]{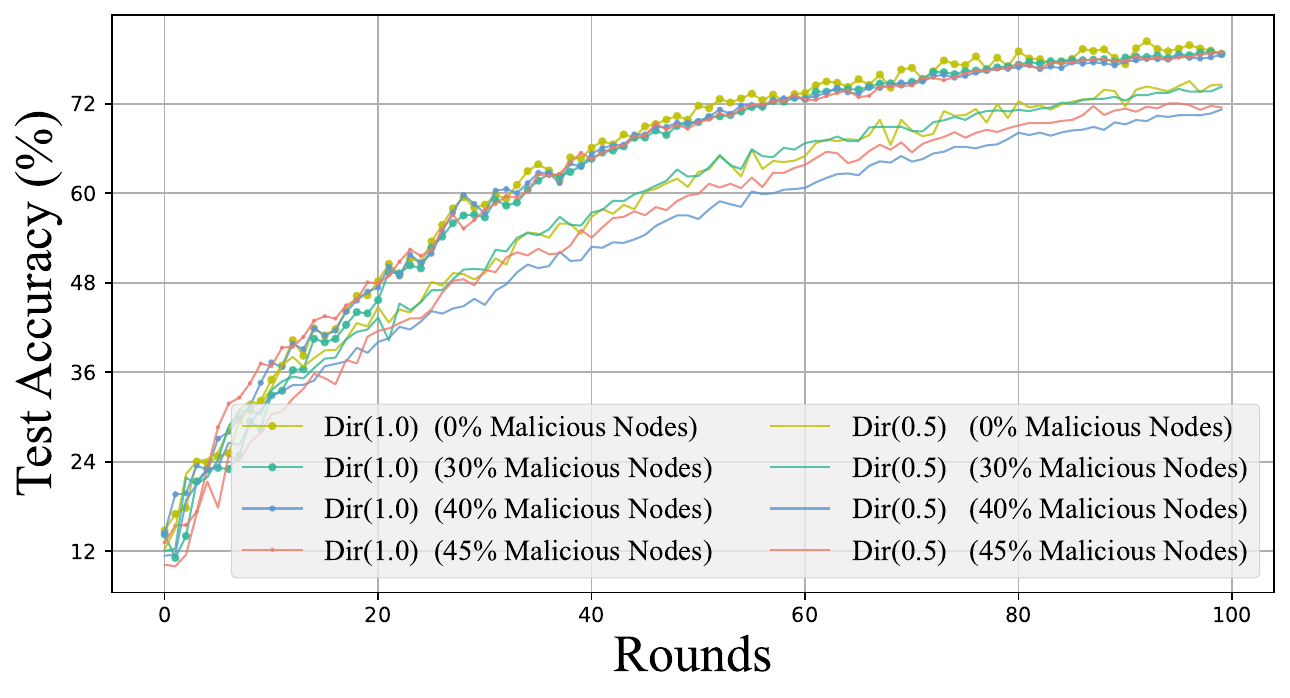}
  }
    \subfloat[DBA Attacks.]
  {
      \label{1121011}  \includegraphics[width=0.45\linewidth]{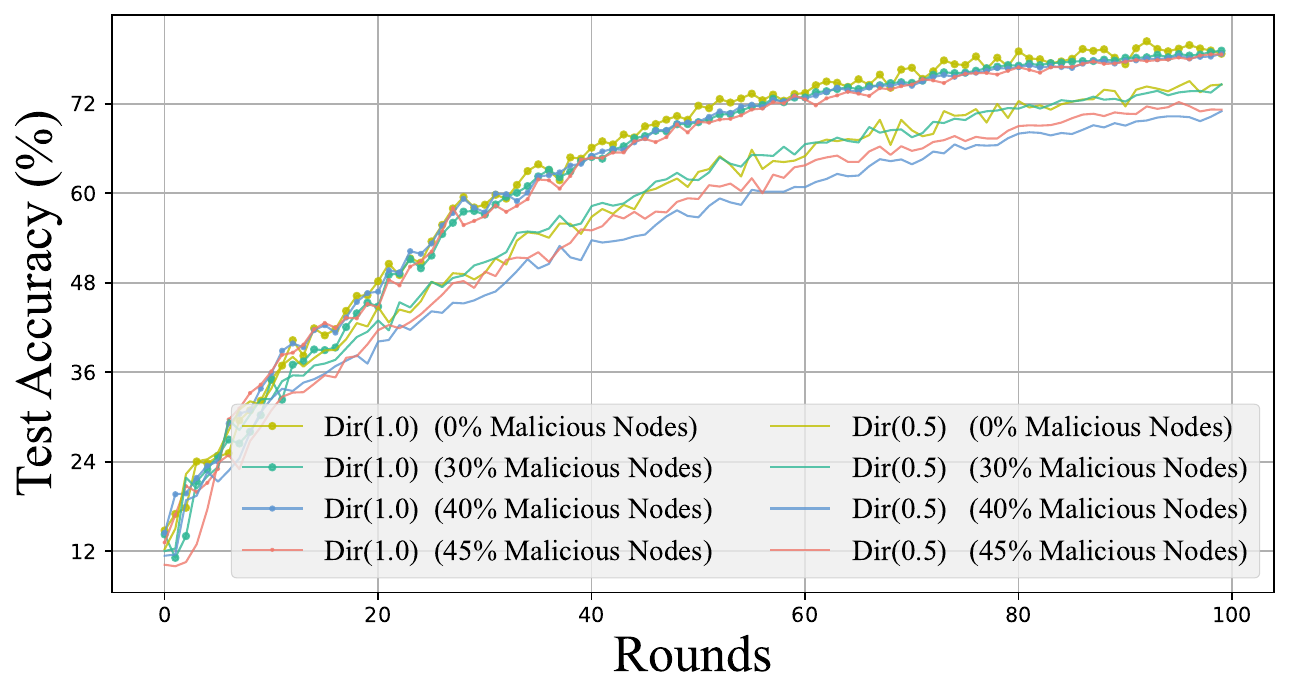}
  }
  
  \subfloat[Neurotoxin Attacks.]
  {
      \label{1121012}  \includegraphics[width=0.45\linewidth]{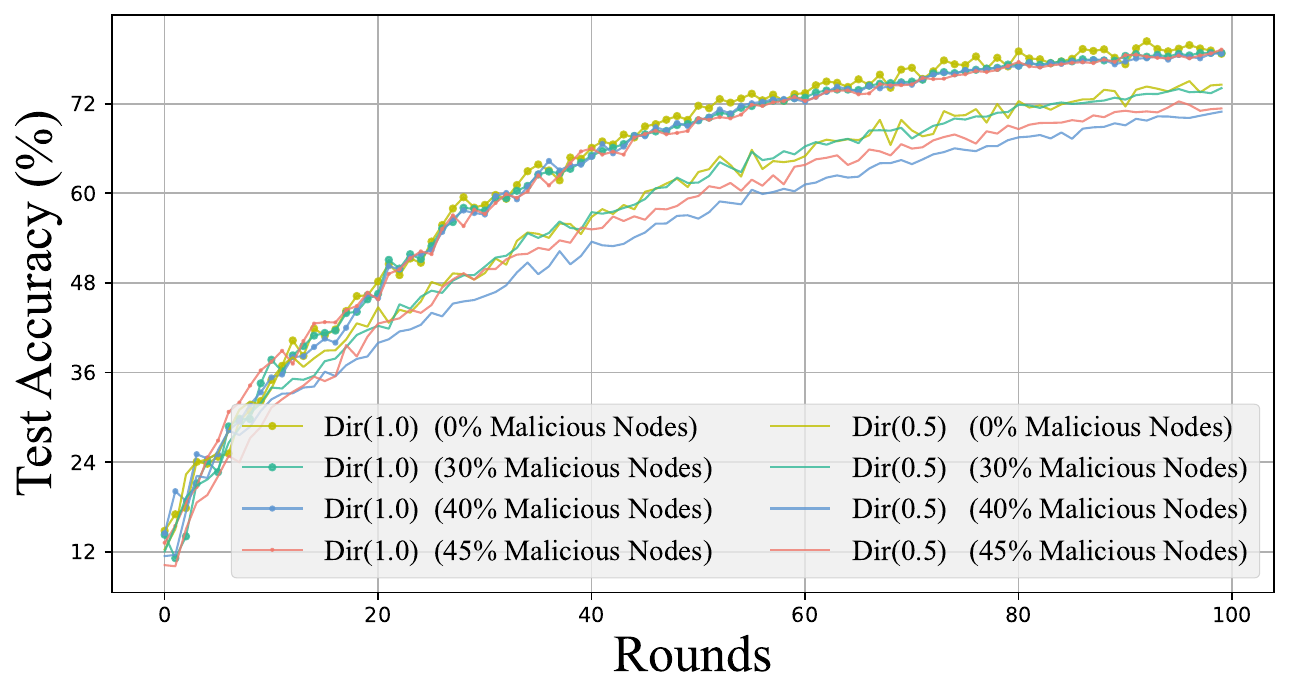}
  }
    \subfloat[PGD  Attacks.]
  {
      \label{1121013}  \includegraphics[width=0.45\linewidth]{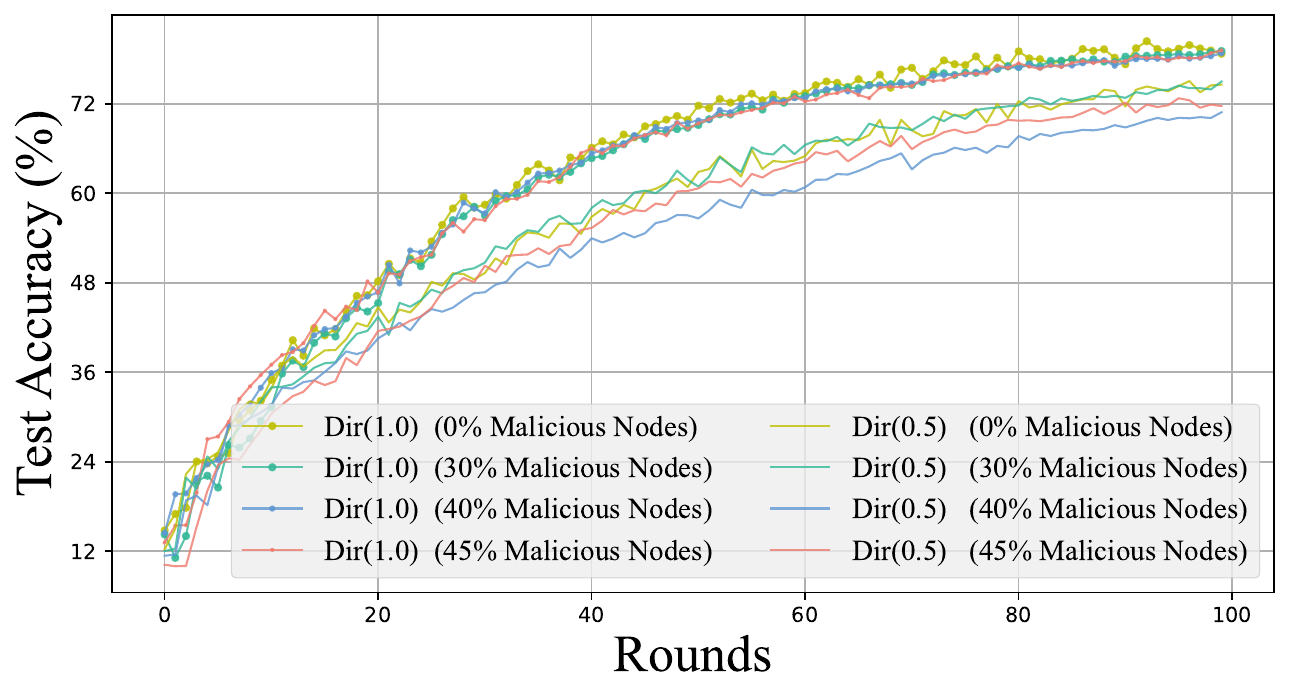}
  }
\caption{Convergence of FedDAB under $\operatorname{Dir}(1.0)$ and $\operatorname{Dir}(0.5)$ settings on the CIFAR10  against various backdoor attacks.}
\label{1121011234}
\end{figure}

To evaluate the convergence performance of FedDAB, we conduct experiments under both the $\operatorname{Dir}(1.0)$ and $\operatorname{Dir}(0.5)$ settings with different proportions of malicious nodes  (i.e., $v \in \{0\%, 30\%, 40\%, 50\%\}$).
As illustrated in Fig. \ref{1121011234}, the convergence curves  under various malicious ratios exhibit trends similar    to that of the no-attack case, indicating that FedDAB does not hinder the convergence of the global model.
In other words,   FedDAB achieves convergence within approximately the same number of training rounds as in the no-attack setting, thereby avoiding additional communication and computation overhead for  the nodes.
Thus, our FedDAB  maintains stable convergence performance across different proportions of malicious nodes.

\subsubsection{Performance of FedDAB under High Proportions of Malicious Nodes}

In FedDAB, a median-based Z-score is employed to filter out local updates exhibiting anomalous $\operatorname{DSS}$ or $\operatorname{SAS}$ values.
Thus, the effectiveness of this median-based scoring mechanism relies on the assumption that benign clients constitute the majority. However, this assumption may not always hold in real-world FL systems.
Therefore, it is necessary to   evaluate FedDAB's defensive performance when the proportion of malicious nodes  exceeds 50\%.
To this end, we set the proportion of malicious nodes to 50\%, 60\%, and 70\%,    using 40\% as the reference setting.
As shown in Table \ref{0121021}, when the proportion of malicious nodes increases from 40\% to 50\%, FedDAB achieves an ASR value that is close to the baseline, indicating that FedDAB does not lose its defensive performance even with 50\% malicious nodes.
Specifically, the model-contrastive term  enhances the consistency of benign updates, causing the DSS and SAS values of benign nodes to cluster more tightly. 
In other words, as long as a malicious update exhibits a significant offset  in either DSS or SAS, FedDAB can identify it as suspicious and exclude it from the aggregation. Thus, even when malicious nodes account for 50\% of all participating nodes, FedDAB remains effective.
When the proportion of malicious nodes further increases to 70\%, the ASR of FedDAB decreases significantly. 
This is because an excessively high proportion of malicious nodes causes malicious updates to dominate the aggregation process, shifting the medians of the DSS and SAS values toward those of malicious updates.
Consequently, FedDAB may struggle to accurately distinguish between malicious and benign updates, resulting in some malicious updates being aggregated while some benign updates are erroneously excluded.
These results reveal the limitations of FedDAB when malicious nodes constitute an overwhelming majority.

\begin{table}[t]
\centering
\caption{Attack success rates of FedDAB under various attacks and high proportions of malicious nodes.}
\scalebox{0.8}{
\renewcommand{\arraystretch}{0.8}
\tabcolsep=0.04cm
\begin{tabular}{cc|cc|cc|cc|c}
\midrule\midrule
\multirow{3}{*}{\makecell{\textbf{Attack}\\\textbf{Type}}}
& \multirow{3}{*}{$\boldsymbol{v}$}
& \multicolumn{6}{c|}{\textbf{Attack Success Rate (\%)$\downarrow$}}
& \multirow{3}{*}{\makecell{\textbf{Avg.}\\\textbf{ASR (\%)$\downarrow$}}} \\

& & \multicolumn{2}{c|}{\textbf{FMNIST}}
  & \multicolumn{2}{c|}{\textbf{CIFAR10}}
  & \multicolumn{2}{c|}{\textbf{CIFAR100}}
  & \\

& & $\operatorname{Dir}(1.0)$
  & $\operatorname{Dir}(0.5)$
  & $\operatorname{Dir}(1.0)$
  & $\operatorname{Dir}(0.5)$
  & $\operatorname{Dir}(1.0)$
  & $\operatorname{Dir}(0.5)$
  & \\
\cmidrule{1-9}

\multirow{4}{*}{Neurotoxin}
& 40\% & 2.86 & 4.79 & 3.02& 3.84 &0.29 & 0.98 & 2.63  \\
& 50\% &  2.92 & 3.07 & 4.83 & 3.95 & 4.61 & 0.47& 3.31 \\
& 60\% &5.58 & 7.91& 6.97 & 8.09 & 5.22 & 4.80 & 6.43  \\
& 70\% &13.25 & 15.38 & 8.03 & 23.94 & 14.26 & 7.41 &13.71  \\
\cmidrule{1-9}

\multirow{4}{*}{PGD}
& 40\% & 2.46 &4.42 & 2.93 & 4.24& 0.23 & 0.75 & 2.51 \\
& 50\% & 4.39 & 6.71 & 5.02 &2.44 & 3.83 & 0.96 & 3.89  \\
& 60\% &  4.34 & 5.82& 7.29 & 5.25& 4.28 & 2.44& 4.90 \\
& 70\% & 23.80 & 15.46 & 12.26 & 10.68 & 16.49 & 5.98& 14.11  \\
\midrule\midrule
\end{tabular}}
\label{0121021}
\end{table}

\subsubsection{Consistency of FedDAB in Non-IID Settings}

To evaluate the effectiveness of our designed model-contrastive term in enhancing the consistency of local updates, we measure the  deviations among local updates in both   magnitude and direction.
The magnitude deviation and directional deviation are defined as follows.
\begin{myDef}\label{11281436} (Magnitude Deviation)
Given that in the $t$-th  round, each benign node $k \in \mathcal{K}_{\mathcal{B}}$ submits its local  update $\Delta_k^t$, the magnitude deviation among local updates  is given by:
\begin{equation}\scalebox{1}{$
\mathrm{Dev}^t_{M} = \log(\sum_{k \in \mathcal{K}_{\mathcal{B}}} \Vert \Delta_k^t - \bar{\Delta}^t \Vert^2),\label{83121261231}$} \nonumber
\end{equation} 
where $\bar{\Delta}^t$ denotes the mean of all local updates,  and $\mathrm{Dev}^t_{M}$ is the magnitude deviation among local updates in the $t$-th round.  
\end{myDef}
\begin{myDef}\label{11281514} (Directional Deviation)
Given that in the $t$-th  round, each benign node $k \in \mathcal{K}_{\mathcal{B}}$ submits its local  update $\Delta_k^t$, the direction deviation among local updates  is given by:
\begin{equation}\scalebox{1}{$
\mathrm{Dev}^t_{D} = \sum_{k \in \mathcal{K}_\mathcal{B}} (1-\operatorname{cos}(\Delta_k^t, \bar{\Delta}^t)),\label{83121262}$}\nonumber
\end{equation} 
where $\operatorname{cos}(\cdot)$ is the cosine similarity between two vectors, and $\mathrm{Dev}^t_{D}$ measures the directional deviation among local updates in the $t$-th round.  
\end{myDef}
\begin{Remark}
If all $k_{1}, k_{2} \in  \mathcal{K}_{\mathcal{B}}$ satisfy $\Delta_{k_{1}}^t = \Delta_{k_{2}}^t$, then the relation  $\bar{\Delta}^t = \Delta_k^t$ holds.
In this case, the local updates are perfectly consistent, which implies  $\mathrm{Dev}^t_{M} = \mathrm{Dev}^t_{D} = 0$.
Conversely, a larger $\mathrm{Dev}^t_{M}$ indicates a greater deviation in update magnitudes, while a larger $\mathrm{Dev}^t_{D}$ reflects significant inconsistency in update directions.
\end{Remark}

\begin{figure}[t]
  \centering
  \subfloat[CIFAR100, $\operatorname{Dir}(1.0)$.]
  {
      \label{112214401}  \includegraphics[width=0.45\linewidth]{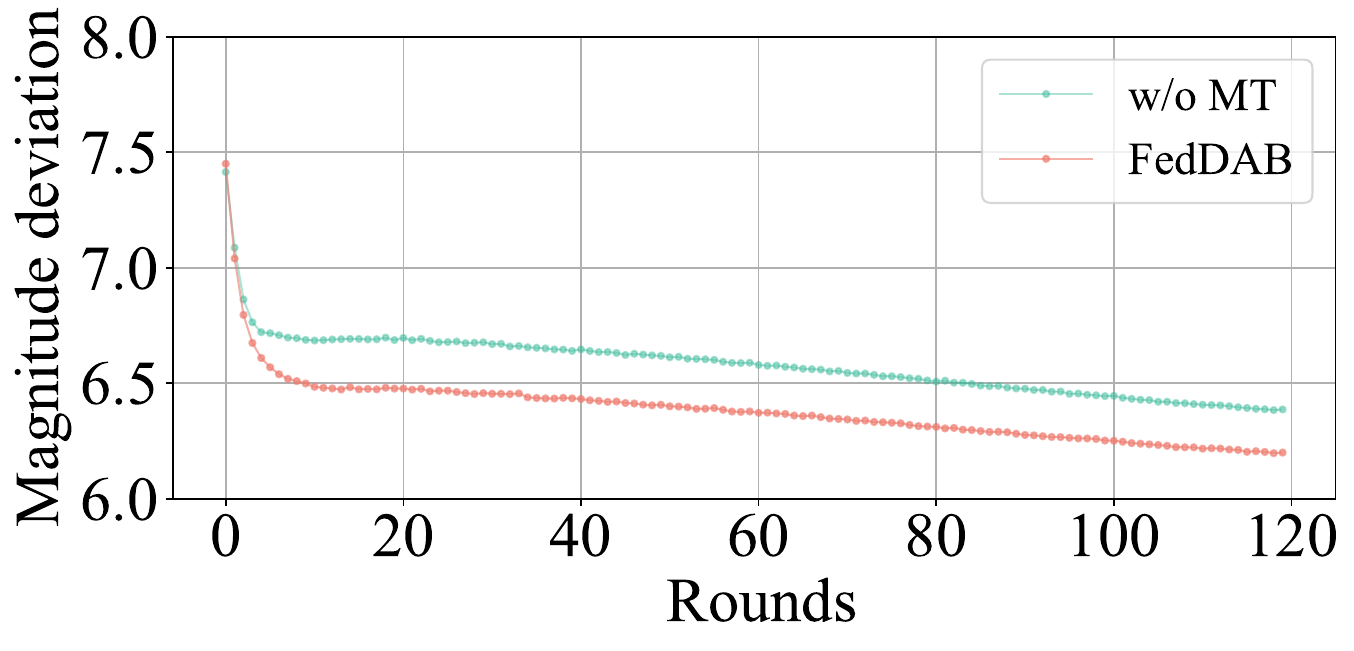}
  }
  \subfloat[CIFAR100, $\operatorname{Dir}(0.5)$.]
  {
      \label{112214402}  \includegraphics[width=0.45\linewidth]{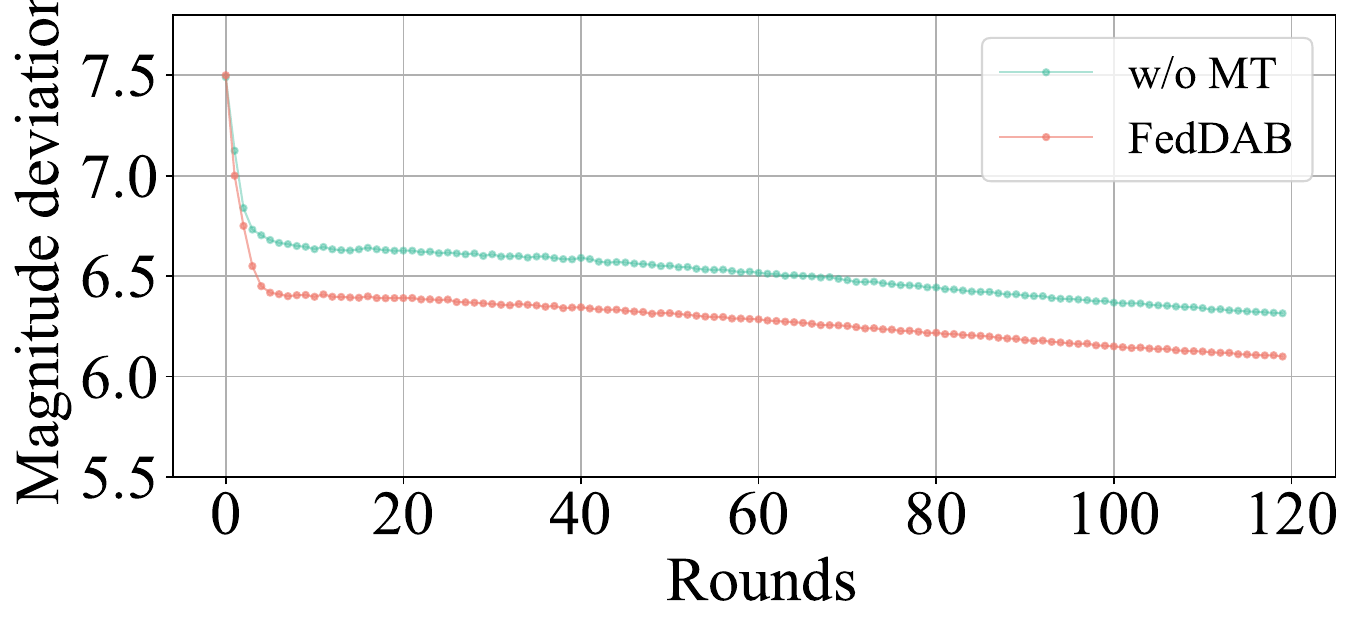}
  }
  
      \subfloat[CIFAR100, $\operatorname{Dir}(1.0)$.]
  {
      \label{112214403}  \includegraphics[width=0.45\linewidth]{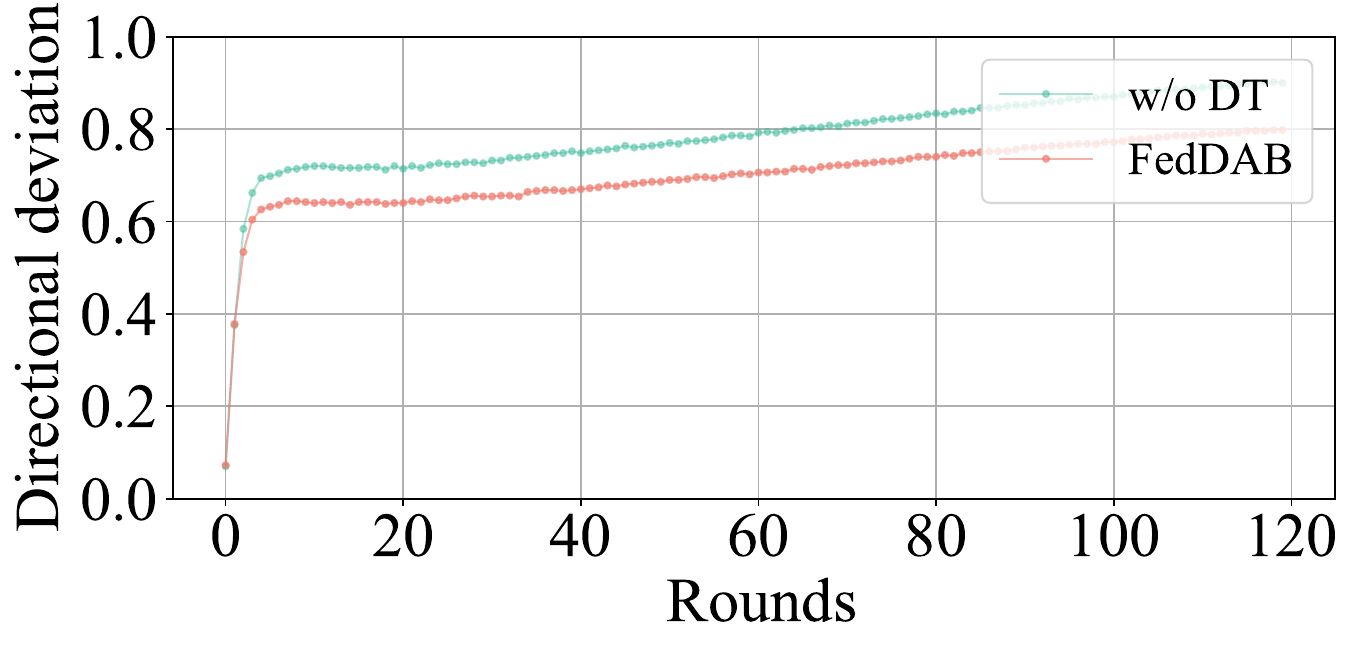}
  }
    \subfloat[CIFAR100, $\operatorname{Dir}(0.5)$.]
  {
      \label{112214404}  \includegraphics[width=0.45\linewidth]{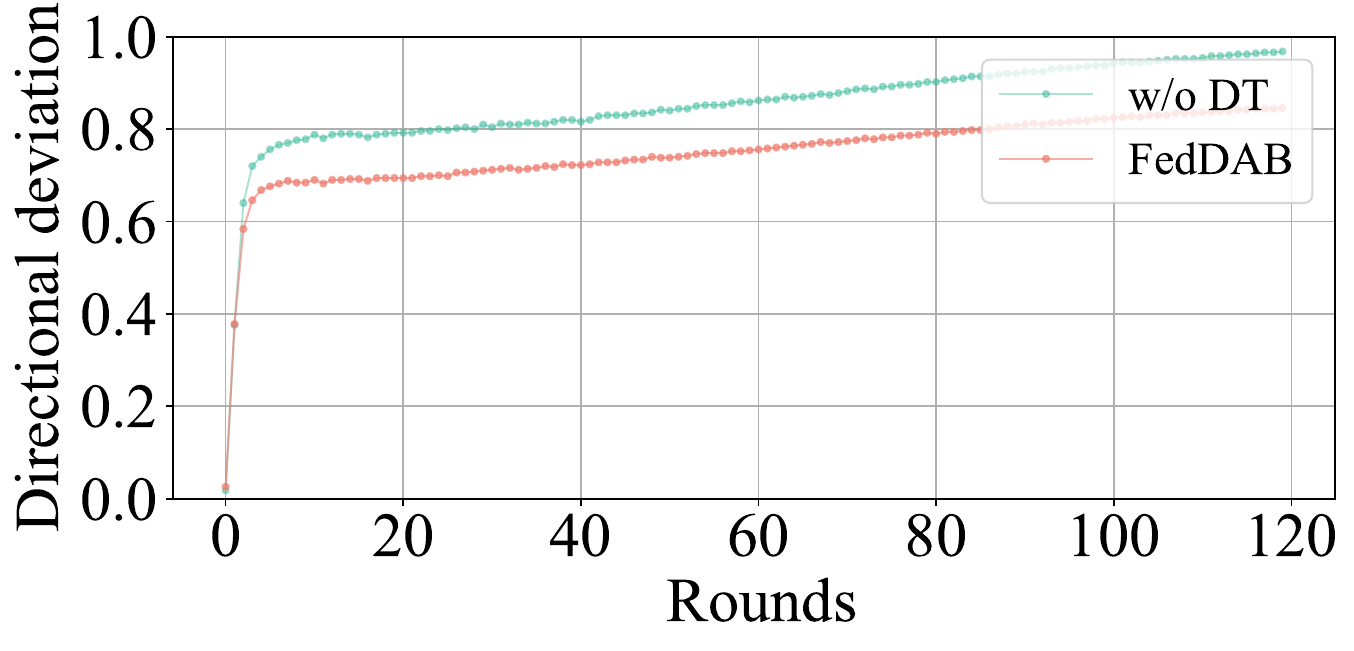}
  }
\caption{Comparison of magnitude deviation and directional deviation with  CIFAR100  under both $\operatorname{Dir}(1.0)$ and $\operatorname{Dir}(0.5)$ settings.}
\label{11221440}
\end{figure}

Following Definition \ref{11281436}, we measure the magnitude deviation  among   local updates generated by different nodes under FedDAB in the absence of backdoor attacks.
For comparison, we evaluate an ablated variant of FedDAB in which the magnitude-alignment term is removed from the local objective (w/o MT).
Experiments are conducted on CIFAR100 under $\operatorname{Dir}(1.0)$ and $\operatorname{Dir}(0.5)$ settings, and the results are reported in Fig. \ref{112214401} and Fig. \ref{112214402}.
We observe that the magnitude deviation under FedDAB is lower than that under w/o MT.
This reduction is  attributed to the magnitude-alignment term introduced in FedDAB's local  objective, which improves magnitude alignment between local and global models and reduces the impact of magnitude inconsistency among benign nodes on \textit{parameter-level checking} step.
Similarly, following Definition \ref{11281514}, we evaluate the directional deviation across nodes' local updates, and  compare FedDAB with an ablated variant that removes the direction-alignment term (w/o DT).
As shown in Fig. \ref{112214403} and Fig. \ref{112214404}, FedDAB exhibits a lower directional deviation than w/o DT,  indicating that the direction-alignment term  improves directional consistency among benign updates.

\subsubsection{Impact of  Parameters $\lambda_{DSS}$ and  $\lambda_{SAS}$ on FedDAB}
\begin{figure}[t]
  \centering
  \subfloat[FMNIST, TA $\uparrow$]
  {
      \label{113020071}  \includegraphics[width=0.3\linewidth]{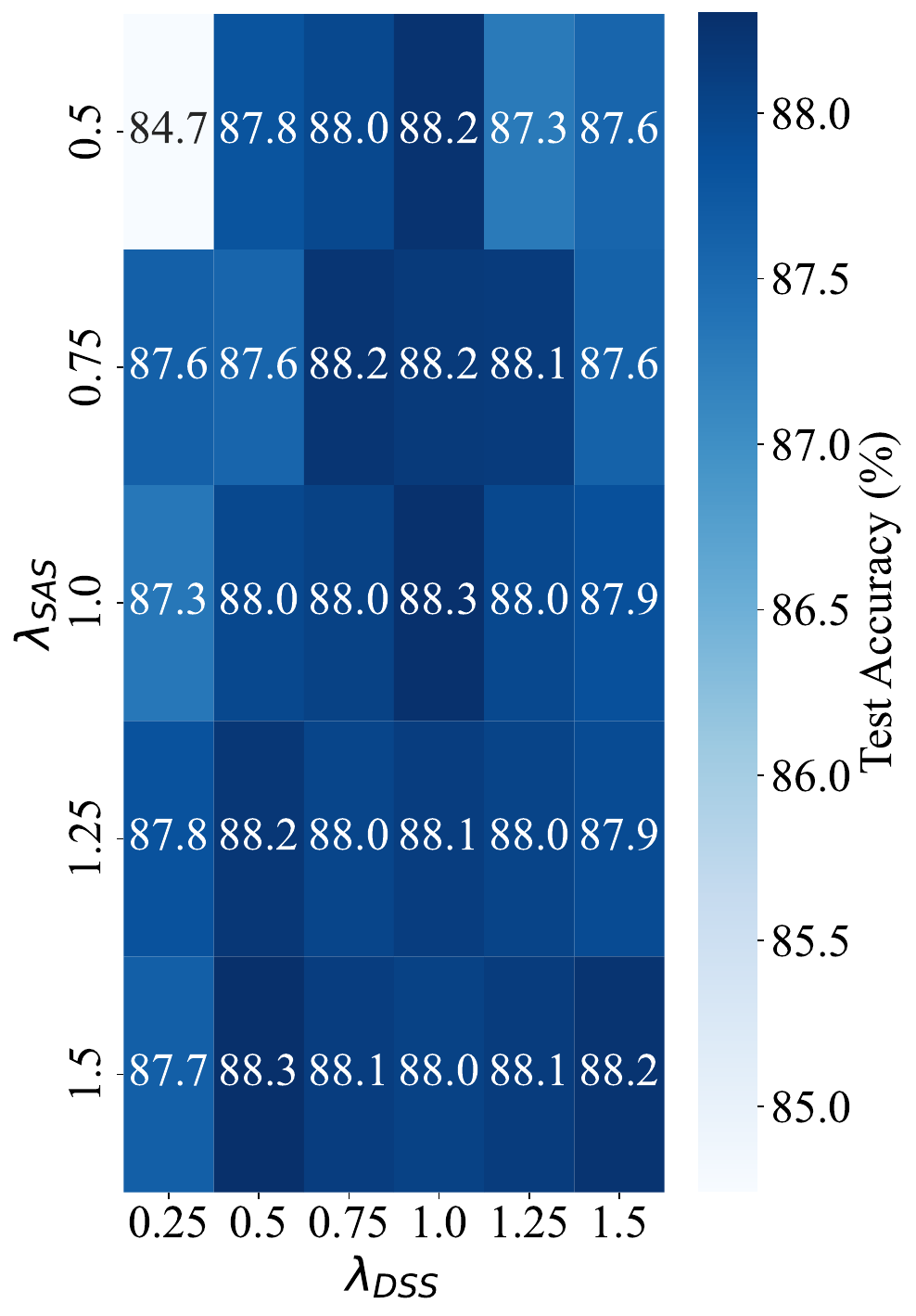}
  }
    \subfloat[CIFAR10, TA $\uparrow$]
  {
      \label{113020073}  \includegraphics[width=0.3\linewidth]{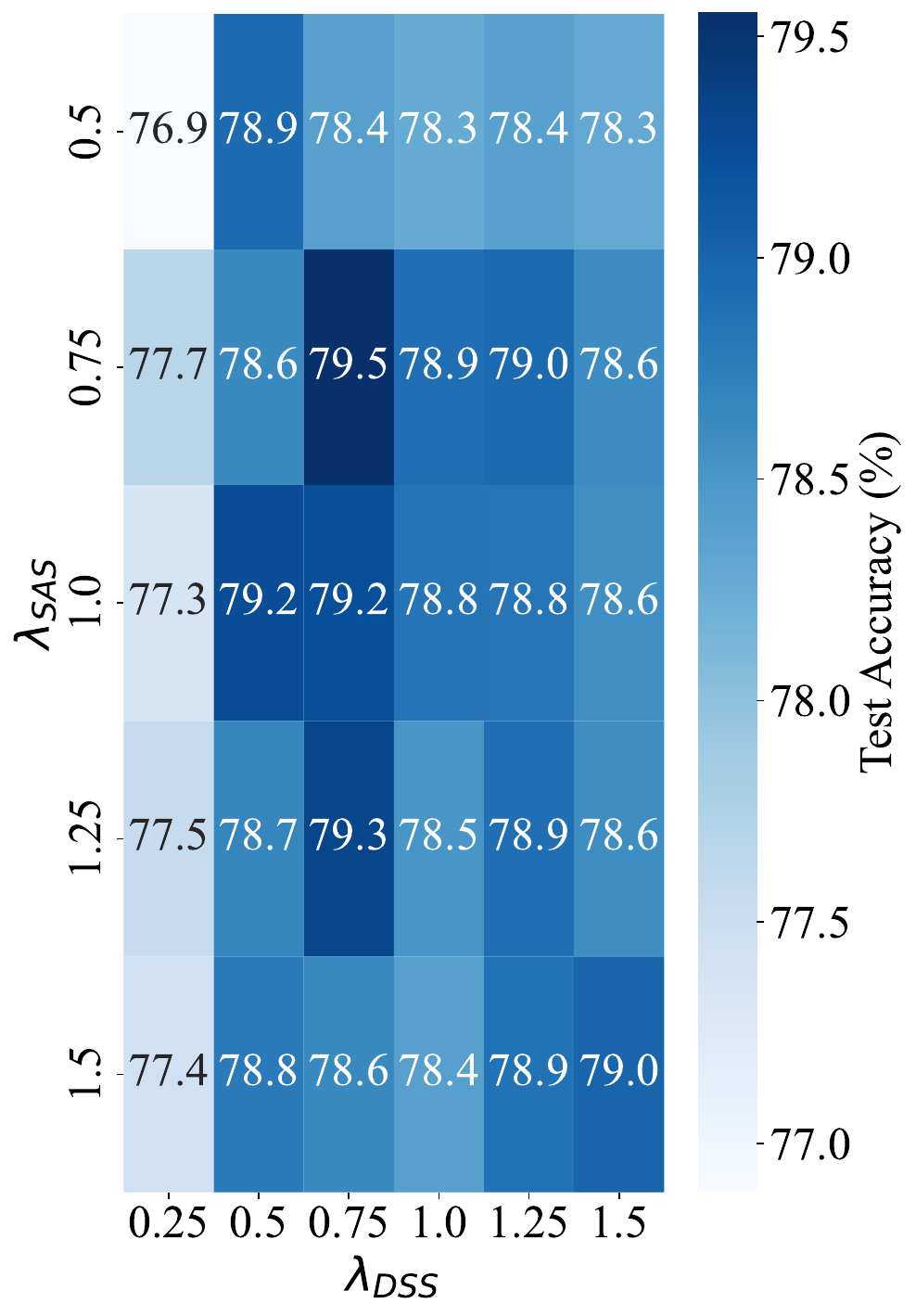}
  }
      \subfloat[CIFAR100, TA $\uparrow$]
  {
      \label{113020075}  \includegraphics[width=0.3\linewidth]{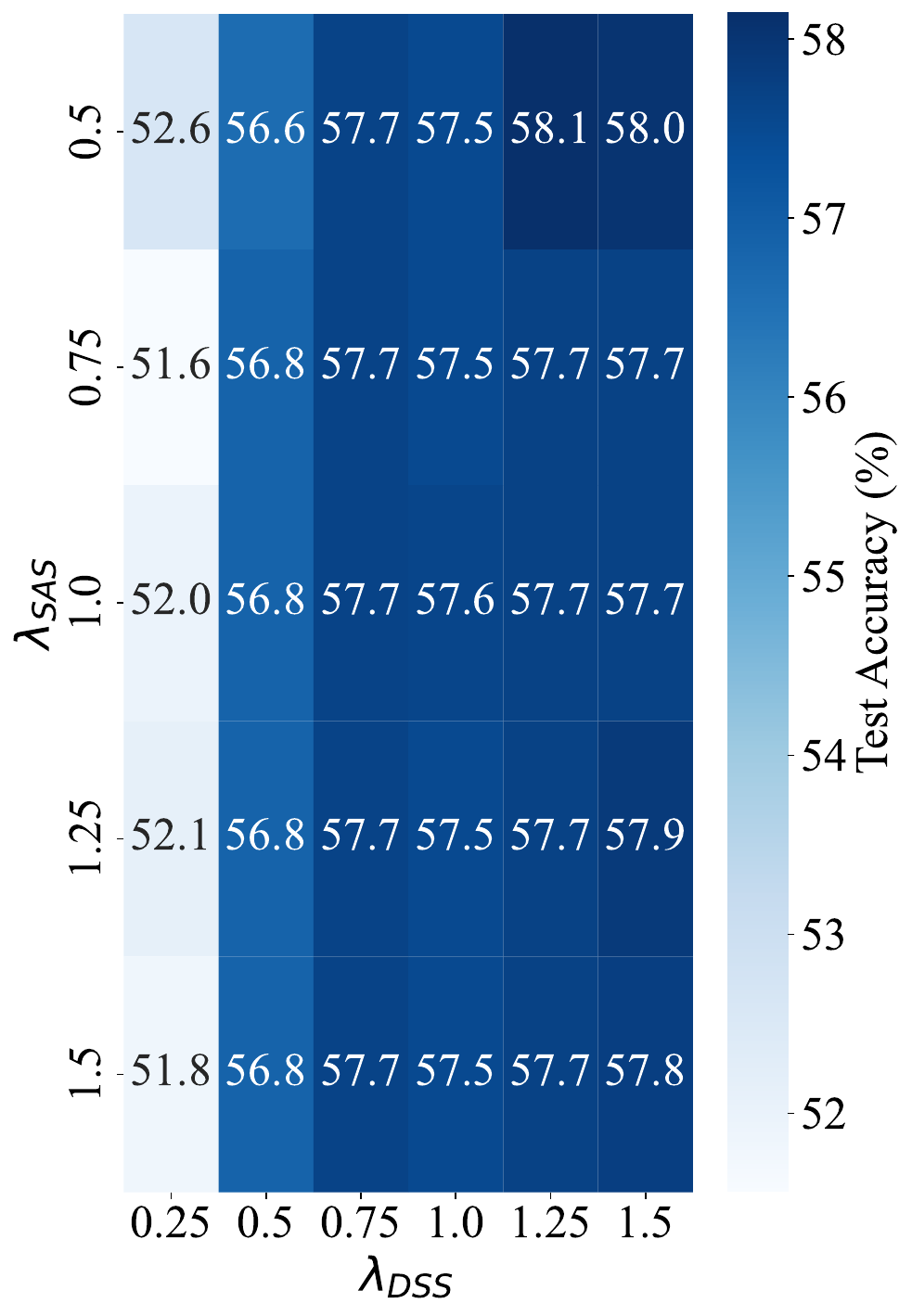}
  } 
  
    \subfloat[FMNIST, ASR $\downarrow$]
  {
      \label{113020072}  \includegraphics[width=0.3\linewidth]{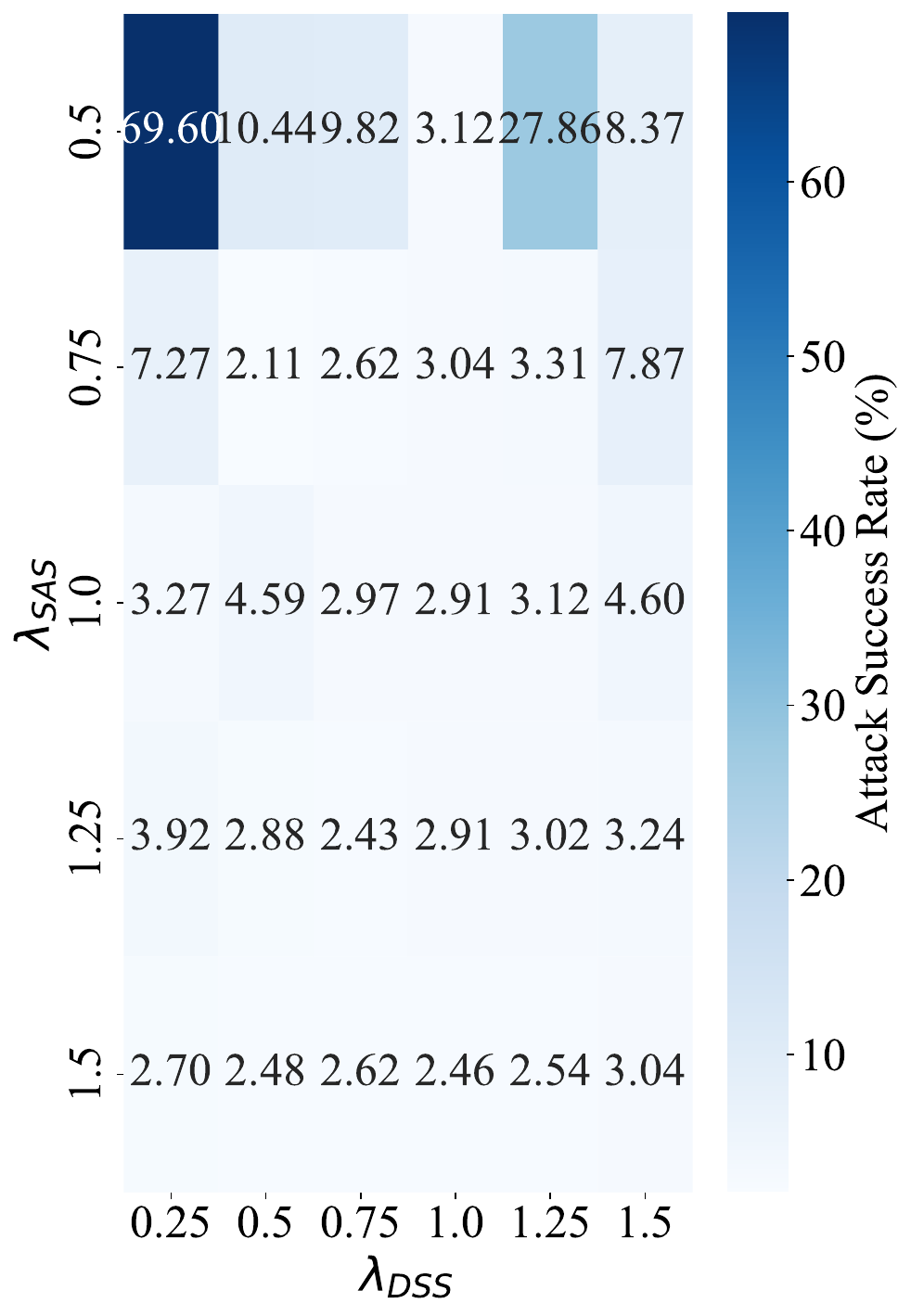}
  }
  \subfloat[CIFAR10, ASR $\downarrow$]
  {
      \label{113020074}  \includegraphics[width=0.3\linewidth]{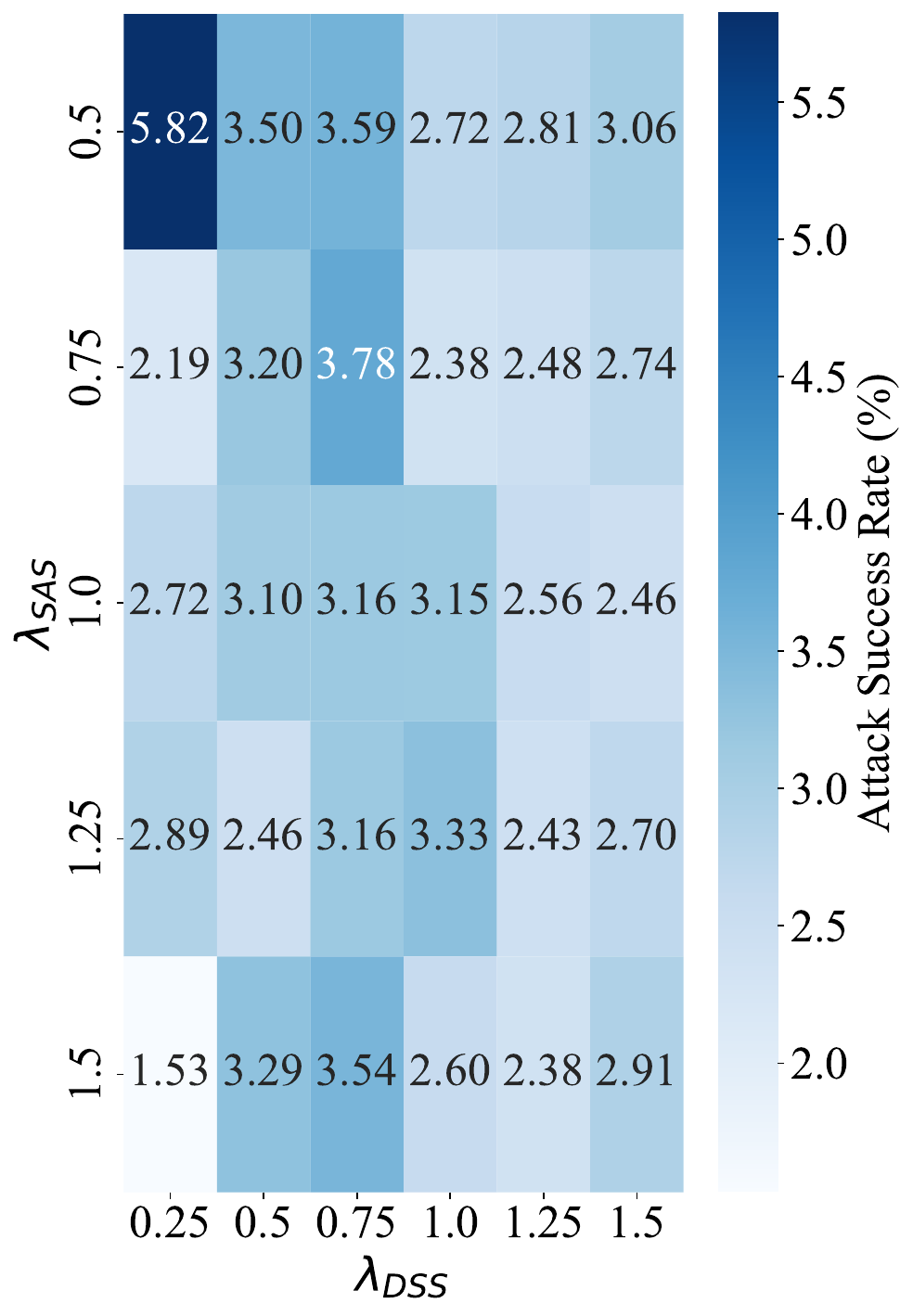}
  }
  \subfloat[CIFAR100, ASR $\downarrow$]
  {
      \label{113020076}  \includegraphics[width=0.3\linewidth]{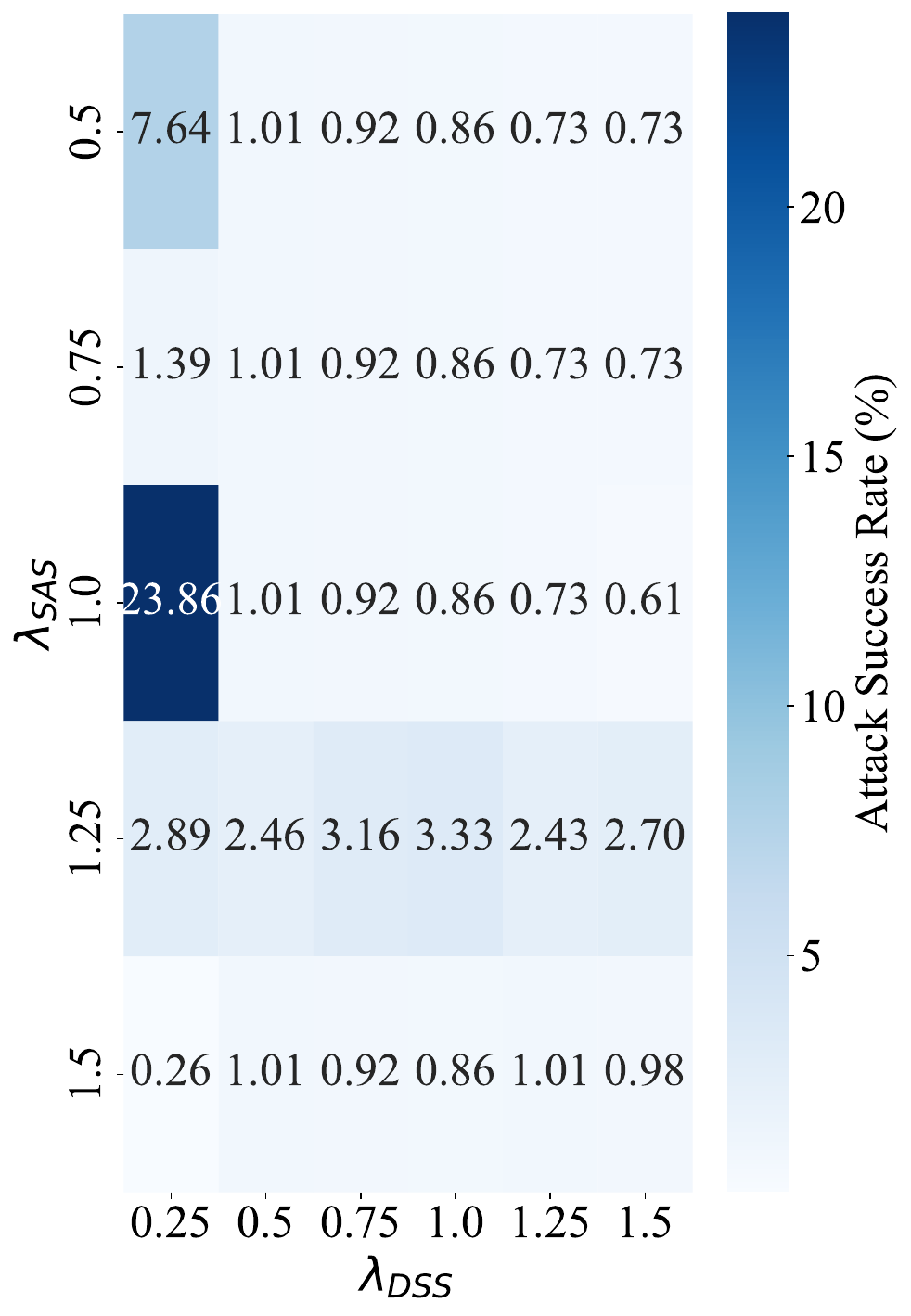}
  }
\caption{Test accuracy and attack success rate of FedDAB with varying  hyper-parameters $\lambda_{DSS}$ and  $\lambda_{SAS}$.}
\label{11302007}
\end{figure}

In FedDAB, the \textit{Anomaly Score Detection} step   computes the median offsets $m_{k,1}^t$ and $m_{k,2}^t$ for each node. 
A node is excluded from the current round if its offset exceeds the corresponding threshold $\lambda_{DSS}$ or $\lambda_{SAS}$.
Thus, a larger threshold allows more nodes to participate in aggregation, while a smaller threshold reduces the number of participating nodes.
To evaluate the FedDAB's sensitivity to  $\lambda_{DSS}$ and $\lambda_{SAS}$, we conduct experiments on  FMNIST, CIFAR10, and CIFAR100. 
Specifically,   $\lambda_{DSS}$ is selected from $\{0.25, 0.5, 0.75, 1.0, 1.25, 1.50\}$, and $\lambda_{SAS}$ is varied over $\{0.5, 0.75, 1.0, 1.25, 1.50\}$.
In the experiments, 30\% of  the nodes launched BadNet attacks, and the data distribution follows $\operatorname{Dir}(1.0)$.
The results are presented in Fig.  \ref{11302007}.
We observe that FedDAB maintains stable    performance across different threshold settings, except when the thresholds are extremely small (e.g., $\lambda_{DSS} =0.25$ or $\lambda_{SAS} = 0.50$).
This is because overly low thresholds filter out a large number of benign nodes, leaving insufficient local updates to compute the global model and  causing the  model to under-fit. 
Thus, to balance defense effectiveness and main-task performance, neither $\lambda_{DSS}$ nor $\lambda_{SAS}$ should be set too low.

\subsubsection{Impact of   Buffer Length $H$ on FedDAB}

\begin{table}[t]
\centering
  \caption{The performance of FedDAB under various attacks with different buffer length $H$. Best rusults are in \textbf{blod}.}
    \scalebox{0.8}{
      \renewcommand{\arraystretch}{0.8}
\tabcolsep=0.054cm
\begin{tabular}{cc|cc|cc|cc|cc}
\midrule\midrule
\multirow{3}{*}{\makecell{\textbf{Attack} \\ \textbf{Types}}} & \multirow{3}{*}{\makecell{\textbf{Buffer}\\ \textbf{Length}}} & \multicolumn{6}{c|}{\textbf{Test Accuracy (\%)$\uparrow$}}    & \multirow{3}{*}[-0.5ex]{\makecell{\textbf{Avg}. \\ \textbf{ASR}$\downarrow$}}    &   \multirow{3}{*}[-0.5ex]{\makecell{\textbf{Avg}. \\ \textbf{TA}$\uparrow$}}                 \\
                      &                                                               & \multicolumn{2}{c|}{FMNIST}  &  \multicolumn{2}{c|}{CIFAR10} &        \multicolumn{2}{c|}{CIFAR100}         &               &           \\
                                                                   &                     & \multicolumn{1}{c}{ASR $(\downarrow)$} & \multicolumn{1}{c|}{TA $(\uparrow)$} & \multicolumn{1}{c}{ASR $(\downarrow)$}  &\multicolumn{1}{c|}{TA $(\uparrow)$}  &\multicolumn{1}{c}{ASR $(\downarrow)$} & \multicolumn{1}{c|}{TA $(\uparrow)$}                                            \\\cmidrule{1-10}   
\multirow{5}{*}{Neurotoxin}    &                                          $H=0$                  & 2.58 & 85.17 & 4.37   &76.73&1.06&  57.37   & 2.67  & 73.09 \\
	                                               &   $H=2$                    & 15.82 & 71.20& 3.68   &78.04&0.85&   \textbf{57.94}   & 6.78 &74.39 \\ 
	                                               &     $H=3$                    & 3.78 & 87.67 & 3.60   &\textbf{78.22}&\textbf{0.71}&    57.75   & 2.70  & \textbf{74.55} \\
                   &    $H=4$                   & \textbf{3.14} & \textbf{87.71} & \textbf{3.42}   &76.88&0.78&    57.47    & \textbf{2.45}  &74.02 \\ 
                                      &    $H=5$                    & 4.54 & 87.26 & 3.64   &78.02 &0.76&   57.59  & 2.98  & 74.29 \\  \cmidrule{1-10}
\multirow{5}{*}{PGD}   &                                          $H=0$                  & 3.42 & 86.28 & 4.09   &77.45&1.04&  57.52  &  2.94  &  73.66\\
	                                               &   $H=2$                    & 2.50 & 87.87& 4.42   &78.81&0.85&   57.48  & 2.59  &74.72 \\ 
	                                               &     $H=3$                    & 2.88 & 88.08 & 3.38   &78.63&0.72&    \textbf{57.65} &    2.33  & 74.79\\
                   &    $H=4$                   & 2.07 & 87.98 & \textbf{3.23}   &\textbf{78.89}&\textbf{0.61}&    57.57    & \textbf{1.97}  & \textbf{74.81}\\ 
                                      &    $H=5$                    &  \textbf{2.04} & \textbf{88.11} & 3.28   &77.44 &0.80&    57.48   & 2.04  &  74.34 \\\midrule\midrule
\end{tabular}}
\label{12041717}
\end{table}

Since benign nodes can  honestly follow the training protocol across all rounds, we incorporate nodes' historical information into the parameter-level checking to better captures their stable behavior.
To evaluate the impact of this mechanism, we set the buffer length to ${0, 2, 3, 4, 5}$, where $H = 0$ indicates that no historical information is used.
The experimental results are reported in Table \ref{12041717}.
We observe that under PGD attacks, compared with $H=0$, FedDAB reduces ASR by more than 1\% on average when $H=4$.
Moreover, FedDAB achieves its highest TA when the buffer length is set to 3 or 4, whereas its TA is the lowest when $H=0$. 
This is because incorporating more historical information improves the robustness  of FedDAB.

\subsubsection{FedDAB on NLP Tasks}
To validate the effectiveness of FedDAB on NLP tasks, we conduct experiments on the Sentiment140 using  ResNet9. 
For the attack setting, poisoned samples are constructed by inserting the  sentence ``This is a backdoor trigger'' into  samples. 
The number of  rounds $T$ is set to 200, while  other hyper-parameters follow the default settings. 
The  results are reported in Table \ref{164164}.
Under different attack ratios, FedDAB achieves ASR values of 39.72\%, 40.03\%, and 45.39\%, corresponding to relative reductions of 7.56\%, 17.21\%, and 40.36\%, respectively, compared with AlignIns.
In addition, FedDAB consistently attains the highest RR, surpassing AlignIns by an average of 15.63\%.
These results verify the robustness of FedDAB on NLP dataset.

\begin{table}[t]
\centering
\caption{Performance under BadNet attacks on Sentiment140.}
\label{164164}
       \renewcommand{\arraystretch}{1}
\resizebox{\columnwidth}{!}{
\tabcolsep=0.094cm
\begin{tabular}{c|ccc|ccc|cc}
\midrule\midrule
\multirow{3}{*}{\textbf{Methods}}
& \multicolumn{8}{c}{\textbf{BadNet Attacks}} \\
\cline{2-9}
& \multicolumn{3}{c|}{\textbf{ASR(\%)}$\downarrow$}
& \multicolumn{3}{c|}{\textbf{RR(\%)}$\uparrow$}
& \multirow{2}{*}[0.5ex]{\textbf{\makecell*[c]{Avg. \\ ASR $\downarrow$}}}
& \multirow{2}{*}[0.5ex]{\textbf{\makecell*[c]{Avg. \\ RR  $\uparrow$}}}\\
& $v$=0.30 & $v$=0.40 & $v$=0.50
& $v$=0.30 & $v$=0.40 & $v$=0.50
& & \\
\hline
MKrum     & 82.48 & 87.23 & 99.88 & 11.46 & 7.24 & 0.17  & 89.86 & 6.29  \\
Foolsgold & 40.45 & 41.06 &  \textbf{44.25} & 52.85 & 51.14 &  \textbf{49.63} & 41.92 &  51.21 \\
RoseAgg   & 63.09 & 67.05 & 79.54 & 29.89 & 29.15 & 24.63 & 69.89 & 27.89 \\
AlignIns  & 42.97 & 48.35 & 76.10 & 51.03 & 48.55 & 9.34  & 55.81 & 36.31 \\
FedDAB    & \textbf{39.72} &  \textbf{40.03} & 45.39 &  \textbf{54.35} &  \textbf{52.63} & 48.83 & \textbf{41.71} & \textbf{51.94} \\
\midrule\midrule
\end{tabular}}
\end{table}

\subsubsection{Time Cost}

\begin{figure}[!t]
  \centering
  \subfloat[Aggregation Time]
  {
      \label{112116462}  \includegraphics[width=0.45\linewidth]{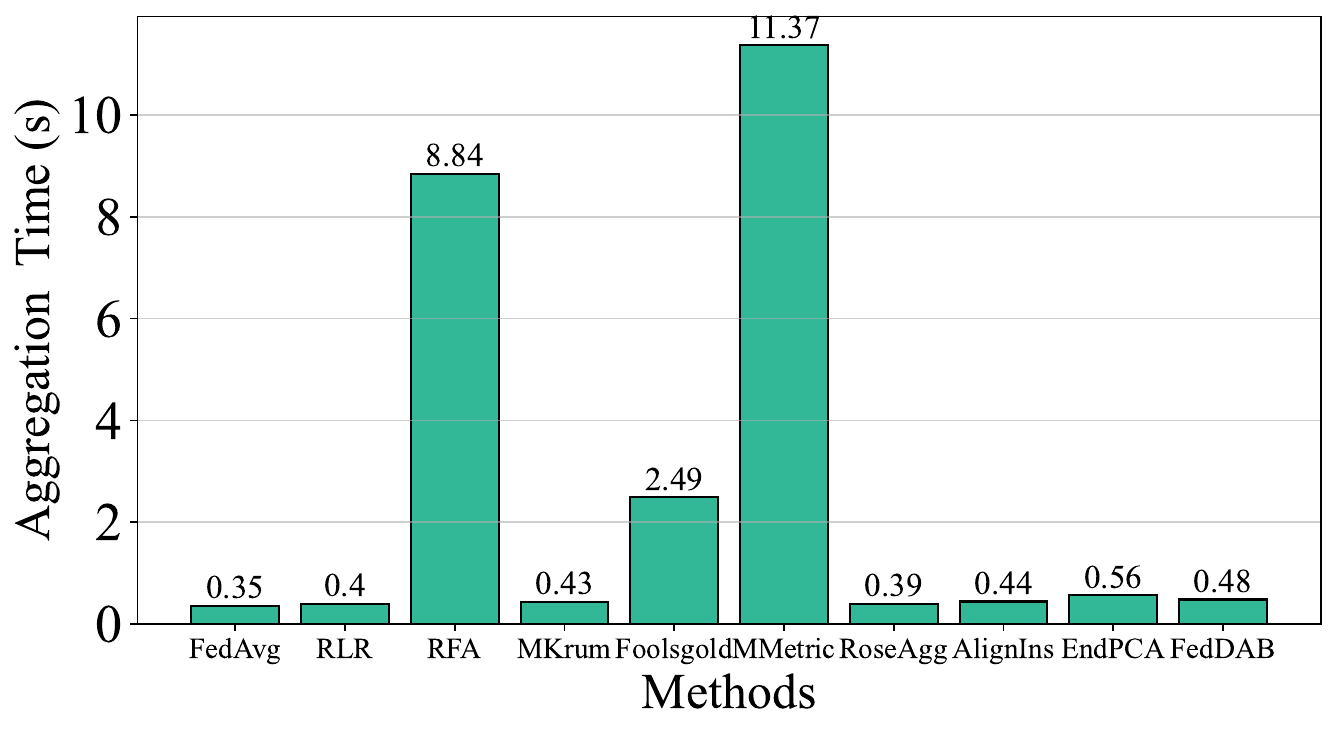}
  }
    \subfloat[Local Training Time]
  {
      \label{112116461}  \includegraphics[width=0.45\linewidth]{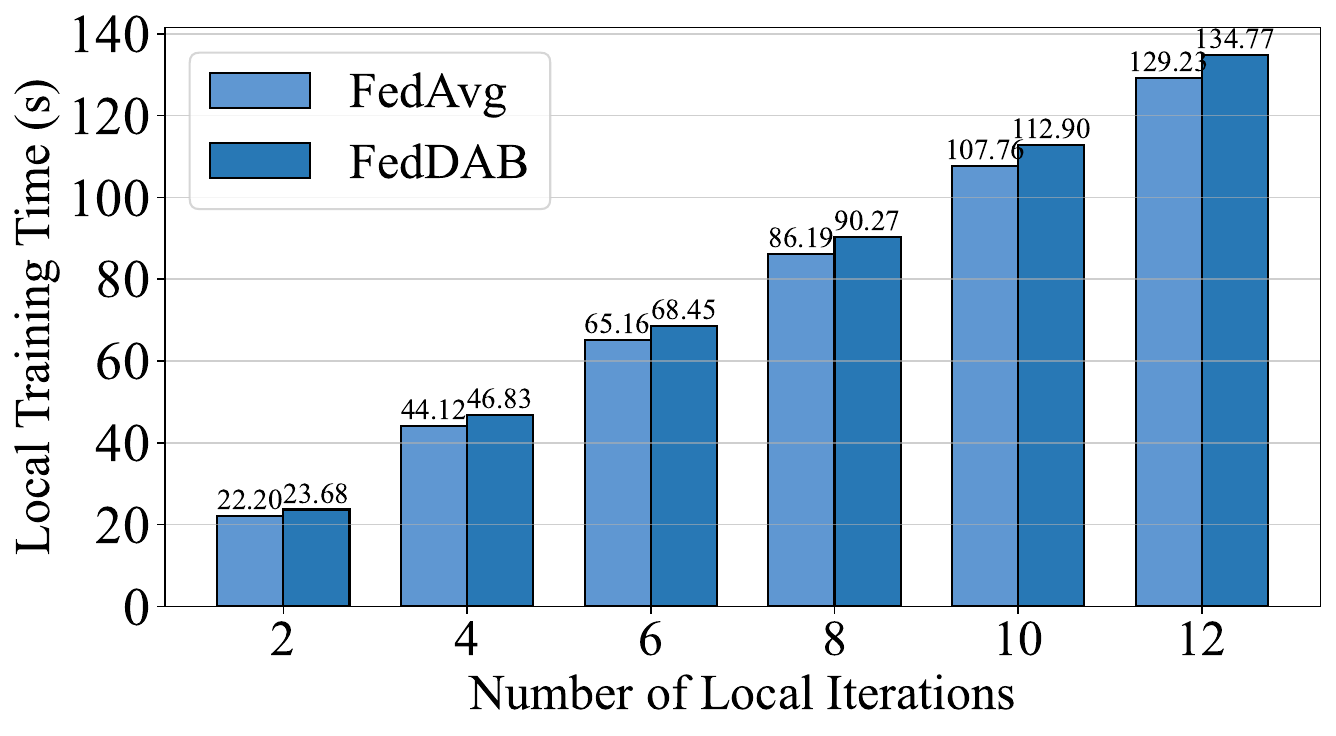}
  }
\caption{Time costs in  FL system.  (a) Aggregation time per round for the defense methods. (b) Training time per round for all nodes. }
\label{11211646}
\end{figure}

To  evaluate the efficiency of FedDAB, we measure both its aggregation time and training time. Specifically, we conduct experiments on the CIFAR10  using the ResNet9 architecture, and compare the aggregation time of FedDAB with that of other defense methods within a single  round. 
All experiments are conducted on an NVIDIA A800 GPU.
The results are presented in Fig. \ref{112116462}.
We observe  that FedAvg achieves the shortest aggregation time because it does not incorporate any defense mechanism. 
In contrast, other defense methods introduce additional detection steps, leading to higher time overhead. 
Notably, the aggregation time for RFA and MMetric is significantly higher than other methods, while  RLR, MKrum, RoseAgg, AlignIns, and FedDAB maintain their aggregation times within the range of 0.4-0.5 seconds. 
This is because the network architecture  typically contains a large number of parameters, leading to substantial computational overhead for RFA when calculating the geometric median of each parameter coordinate during  aggregation. 
Meanwhile, MMetric incurs  significant time costs  by evaluating multiple metrics for each model update.
Overall, FedDAB does not introduce noticeable aggregation overhead, and its time efficiency is comparable to that of most defense methods.
Except for the aggregation time, we  evaluate the local training time of FedDAB. Specifically, we measure the training time of FedDAB and FedAvg under different numbers of local iterations, as shown in Fig. \ref{112116461}. 
We observe that FedDAB's training time is slightly higher than that of FedAvg.
This is because FedDAB introduces a model-contrastive term in its optimization objective to enhance the consistency among benign local updates.
Thus, it is reasonable for FedDAB to incur additional computational cost during local training.

\begin{table}[t]
  \caption{ Memory overhead of   local sign buffer  in FedDAB.} 
\centering
       \renewcommand{\arraystretch}{1}
\resizebox{\columnwidth}{!}{
\begin{tabular}{c|cccccccc}
\midrule\midrule  Number of Nodes &20&30&40&50&60&70&80&100\\\cline{2-9}
Memory overhead  (GB) &1.17&1.76&2.33&3.01&3.54&4.11&4.70&5.90\\
\midrule\midrule\end{tabular}}
\label{84165112}
\end{table}
\subsubsection{Memory overhead}
 FedDAB  requires maintaining  per-node historical buffers  $\mathcal{S}_l$, which introduces additional memory cost.
To evaluate this overhead, we conduct experiments on CIFAR10 with ResNet9 with the historical buffer length set to 3.
As reported in Table \ref{84165112}, the server-side memory overhead grows linearly with the number of nodes.
This overhead is acceptable in edge computing scenarios, as it is incurred only on the server side and does not impose additional storage burden on edge nodes.
Notably, the server side  has more sufficient storage resources than  edge nodes.

\section{Conclusion}
We propose  a two-phase method FedDAB to defend against backdoor attacks in FL systems.
The method combines local contrastive regularization with server-side alignment checking.
Experiments validate the effectiveness of  FedDAB, and show that it outperforms existing defense methods.
However, FedDAB still has certain limitations.
Specifically, malicious nodes may abuse their access to the global model by copying or distributing it without authorization \cite{11143953}.
In future work, we will investigate the integration of intellectual property protection mechanisms into FL systems while preserving their robustness against backdoor attacks.

\bibliographystyle{IEEEtran}
\bibliography{ref}

\end{document}